\documentclass[12pt,preprint]{emulateapj}

\slugcomment{}

\shorttitle{X-ray sources in GOODS-S}
\shortauthors{Cappelluti et al.}

\begin{document}


\title{ Chandra counterparts of CANDELS GOODS-S sources.}


\author{N. Cappelluti\altaffilmark{1,2,3}, A. Comastri\altaffilmark{1}, A. Fontana,\altaffilmark{4}, G. Zamorani\altaffilmark{1}, R. ~Amorin \altaffilmark{4}, M. ~Castellano\altaffilmark{4}, 
 E. ~Merlin\altaffilmark{4},  P. ~Santini\altaffilmark{4}, D. ~Elbaz\altaffilmark{5},  C. ~Schreiber\altaffilmark{5},  X. ~Shu \altaffilmark{5,10}, T. ~Wang\altaffilmark{5,11}, J. ~S. ~Dunlop\altaffilmark{6},  N. ~Bourne\altaffilmark{6}, V. ~A. ~Bruce\altaffilmark{6}, F. ~Buitrago \altaffilmark{6,12,13}, Micha{\l} J.~Micha{\l}owski\altaffilmark{6}, S. ~Derriere\altaffilmark{7}, H.~ C. ~Ferguson\altaffilmark{8}, S. ~M. ~Faber\altaffilmark{9}, F.~Vito\altaffilmark{1,14}}

\altaffiltext{1}{INAF - Osservatorio Astronomico di Bologna, Via Ranzani 1, I - 40127, Bologna, Italy}
\altaffiltext{2}{Department of Physics, Yale University, P.O. Box 208121, New Haven, CT 06520, USA}
\altaffiltext{3}{Yale Center for Astronomy \& Astrophysics, Physics Department, P.O. Box 208120, New Haven, CT 06520, USA}
\altaffiltext{4}{INAF - Osservatorio Astronomico di Roma, Via Frascati 33, I - 00040 Monte Porzio Catone (RM), Italy}
\altaffiltext{5}{Laboratoire AIM-Paris-Saclay, CEA/DSM/Irfu - CNRS - Universit\'e Paris Diderot, CEA-Saclay, pt courrier 131, F-91191 Gif-sur-Yvette, France}
\altaffiltext{6}{SUPA\thanks{Scottish Universities Physics Alliance}, Institute for Astronomy, University of Edinburgh, Royal Observatory, Edinburgh, EH9 3HJ, U.K. }
\altaffiltext{7}{Observatoire astronomique de Strasbourg, UniversitŽ de Strasbourg, CNRS, UMR 7550, 11 rue de lÕUniversitŽ, F-67000 Strasbourg, France}
\altaffiltext{8}{Space Telescope Science Institute, 3700 San Martin Drive, Baltimore, MD 21218, USA}
\altaffiltext{9}{UCO/Lick Observatory, University of California, 1156 High Street, Santa Cruz, CA 95064, USA}
\altaffiltext{10}{Department of Physics, Anhui Normal University, Wuhu, Anhui, 241000, China}
\altaffiltext{11}{School of Astronomy and Astrophysics, Nanjing University, Nanjing, 210093, China}
\altaffiltext{12}{Instituto de Astrof\'{\i}sica e Ci\^{e}ncias do Espa\c{c}o, Universidade de Lisboa, OAL, Tapada da Ajuda, PT1349-018 Lisbon, Portugal}
\altaffiltext{13}{	Departamento de F\`{i}sica, Faculdade de Ci\^{e}ncias, Universidade de Lisboa, Edif\`{i}cio C8, Campo Grande, PT1749-016 Lisbon, Portugal}
\altaffiltext{14}{Department of Astronomy and Astrophysics, The Pennsylvania State University, University Park, PA 16802, USA}

\begin{abstract}

  Improving the capabilities of detecting faint X-ray sources is fundamental to increase the statistics on 
  faint high-z AGN and star-forming galaxies.
   We performed a  simultaneous Maximum Likelihood PSF fit in the [0.5-2] keV and [2-7] keV energy bands of the 4 Ms
{\em Chandra} Deep Field South (CDFS) data   at the position of the 34930 CANDELS H-band selected galaxies. 
 For each detected source we provide X-ray photometry and optical 
counterpart validation.  We validated this technique by means of a raytracing simulation. 
   We detected a total of 698 X-ray point-sources with a likelihood $\mathcal{L}$$>$4.98 (i.e. $>$2.7$\sigma$). We show that  the prior
knowledge of a deep sample of Optical-NIR  galaxies leads to a significant increase of the detection of faint  (i.e. $\sim$10$^{-17}$ cgs in the [0.5-2] keV band) 
sources with respect to "blind" X-ray detections.  By including previous X-ray catalogs, this work increases the total number of X-ray sources detected in 
the 4 Ms CDFS, CANDELS area to  793, which represents the largest sample of extremely faint X-ray sources assembled to date.
  Our results suggest that a large fraction of the optical counterparts of our X-ray sources determined 
by likelihood ratio actually coincides with the priors used  for the source detection.
Most of the new detected sources  are likely star-forming galaxies or faint
absorbed AGN. 
We identified a few sources sources with putative photometric
redshift z$>$4. Despite the low number statistics and the uncertainties on the photo-z, this sample significantly
increases the number of X--ray selected candidate high-z AGN. 

\end{abstract}

\keywords{galaxies: active--- galaxies: active, (galaxies:) quasars: supermassive black holes,
 galaxies: high-redshift } 



{\it Facilities:}  \facility{HST (WFC3)}, \facility{CXO (ACIS)}.

%
\section{Introduction}

The scientific return of deep X--ray surveys is maximized in those
regions of the sky intensively covered by longer wavelength
observations.  For example, the study of the accretion and star
formation processes and their cosmic evolution is routinely performed
combining observations obtained in the X--ray and in the optical and near
infrared bands.
It is widely accepted that all bulged galaxies host a  Super Massive Black Hole (SMBH) 
in their center and a fraction of them, roughly of the order of a few percent, show some kind
of nuclear activity. Luminous X--ray emission is a clear signature of nuclear activity 
produced in the vicinity of the central black hole (BH). Also non--active galaxies 
emit X--ray light, at luminosities much lower than that produced by AGN, due to stellar driven processes such as accretion
onto binaries and supernovae remnants.
As a consequence,  X--ray luminosity is also a probe of the Star Formation Rate \citep[SFR,][]{fab,ran,min,basu}. 
Together with Clusters of Galaxies, Active Galactic Nuclei (AGN)  and Star-forming galaxies (SFG) are the 
three main ingredients of the extragalactic Cosmic X-ray Background
(CXB).
 {\em Chandra} and XMM-{\em Newton}  
surveys were able to resolve  a large fraction  of the extragalactic CXB in discrete sources. 
The yet unresolved fraction is thought to be made by a mix of faint SFG at
moderate to high redshifts and low luminosity AGN. 


The selection of sizable samples of faint AGN is fundamental to
understand AGN evolution  and to constrain models of SMBH formation
especially at  high--z.  
So far X--ray surveys have  sampled  the bright end  (L$_X \ge$10$^{43}$)  of the AGN 
X--ray Luminosity Function (XLF) up to z$\sim$5 \citep[see e.g.][]{ueda,has,aird,ebrero,miy15,vito}. 
At higher redshifts only  a handful of  very bright AGN powered by
massive BH are known, but the low luminosity tail of the XLF remains
unknown.
These "missing"  black holes are the key to  understand the mass 
build-up of SMBH in the first Gyr of the Universe and to improve our
understanding of their formation and early evolution.
In fact, the mechanism of SMBH  formation is still a matter of debate since their growth up to 
$\sim$10$^{9}$ M$_{\odot}$   by z$\sim$7 \citep{mort} cannot 
be explained by Eddington limited accretion onto ordinary stellar remnant seed black holes in such a short time. 
This {\em problem} can be solved by invoking the formation of massive BH  seeds at z$\ge$10 or
 supercritical accretion episodes \citep{mad}. \\

Theorists are debating if the SMBH seeds were formed by the collapse
of an early generation of stars (named Population III, POPIII)  or from the direct collapse
of pristine gas clouds (Direct Collapse Black Holes, DCBHs). 
The end point of the evolution of a POPIII star  is a $\sim$10$^{1-2}$ M$_\odot$ BH, while
DCBH can  easily reach $\sim$10$^{5-6}$  M$_\odot$ already at z$\ge$10 \citep{yue13}.   
\citet{vol10} predicts that, if the main SMBH seeding mechanism was DCBH, then 
the number density of low luminosity AGN should rapidly decline at z$\ge$3, while if the seeding
mechanism was mainly due to POPIII stars then the number density of low luminosity AGN at z$\ge$3 
should decline more gently.  
Unfortunately there are no direct observational evidences of SMBH seeds,
though indirect arguments based on the X--ray and near--IR backgrounds 
\citep[see e.g.][]{kash12,cap13,yue13} or stacking \citep{tre} suggest
that significant progresses may be obtained by a synergic
multi--wavelength approach. 

By combining {\it Chandra} 2 Ms deep X--ray observations (Luo et
al. 2008) and optical/ near--infrared images in the z,K,IRAC images in
the GOODS-MUSIC field along with F160W data in the ERS \citep[Early
Release Science,][]{graz} region Fiore et al. (2012) pushed the formal
detection limits of the X--ray images at deeper levels using the
optical near infrared images as priors. Giallongo et al. (2015)
improved the method outlined above using 4 Ms {\it Chandra} data and
F160W GOODS images. The optical/near infrared priors have then been
used to select high redshift ($z>$ 4) AGN and evaluate their impact on
the reionization history of the Universe \citep{gial}.  Pushing the
limits of deep {\em Chandra} Surveys towards ultra faint fluxes would
also allow to boost the detections of faint \citep{leh} {\it normal}
(SFG) galaxies which start to outnumber AGN around 10$^{-17}$ erg
cm$^{-2}$ s$^{-1}$ in the 0.5--2 keV band.  The detection of
additional very faint X-ray sources and their identification in the
optical/NIR may lead to the discovery of moderate redshift
(z$\sim$1-2) SFG and improve the current knowledge of the cosmic
evolution of binaries in galaxies.  The evolution of SFGs has been
mostly determined via stacking of optically selected samples
\citep{basu}. Stacking is a powerful tool, however the outcomes of
these investigations are strongly influenced by the choice of the
reference sample.  Samples of X-ray detected SFGs are available only up
to z$\sim$1.3 \citep{min14} making it difficult to perform a direct
determination of their evolution around and beyond the peak of cosmic
star formation at z$\sim$2-3. In order to increase these sample sizes
we need to boost our efficiency in detecting faint sources by
developing new source detection techniques.

Unfortunately, the sky area sensitive to extremely faint fluxes (and
luminosities) is very small and therefore only a handful
of faint sources (either high--z AGN or SFG) have been detected so far. 
While we cannot push the flux limit to fainter fluxes,  we can develop methods
that allow us to increase the efficiency of source detections.

  The method described in this paper is conceptually similar to that followed by Giallongo et al. (2015) and originally proposed
 in Fiore et al. (2012), but differs from standard methods usually adopted in the literature. 
 The most recent and comprehensive discussion is reported in \citet{hsu} where the optical/NIR 
 counterparts are searched within the X-ray positional error box.
The here proposed method maximizes the number of CANDELS sources with an X-ray counterpart. 
The advantage  here is that, thanks to the unprecedented depth of WFC3 images (down to m$_{AB}\sim$29-30 in H-band),
 almost the totality  of the counterparts  of the X-ray sources are already detected in the CANDELS H-band catalogue. 
 In fact the likelihood that a {\em Chandra}  source has a counterpart
with H magnitude below the detection limit of WFC-3 is very low. Moreover, 
in this paper we  take advantage of  the superb {\em Chandra} angular
resolution and astrometric accuracy, that guarantees the capability of associating a very large fraction of 
X-ray sources to optical/NIR counterparts in HST images
\citep{xue,civ10,hsu}. 
As mentioned above a well established method in the literature, is to assign 
a counterpart to the X-ray detection with the Likelihood Ratio (LR) technique \citep[see e.g.][]{cil,bru07,civ10}.
Here we employ the LR technique to evaluate the reliability of our source detection, counterpart assignment and to complement
our catalog in the few cases where our method fails. Other authors used a similar approach but validating the associations with a 
 bayesian analysis \citep[e.g.][]{hsu}.
The CDFS/GOODS-S was observed by  HST-WFC3/ACS in 
the Cosmic Assembly Near-Infrared Extragalactic Legacy Survey (CANDELS) which incorporates 
a wide 0.048 deg$^{2}$ observation plus the so-called Hubble Ultradeep Field (UDF) and, thanks to
the extraordinary sensitivity, reaches  H-Band depth of m$_{AB}\simeq $28 \citep{guo13}.\\
The outstanding quality of the HST CANDELS catalog, combined 
with the sub-arcsec angular {\em Chandra} resolution, makes it possible to directly
perform a PSF fitting of X-ray data at the position of each HST source.  

The overall approach is similar to that
pioneered by Fiore et al. (2012), but it benefits of improved detection
techniques and homogenoeus treatement of the data as well as of
extensive simulations.
Even though, at the time of writing, a large fraction of the ultradeep 7 
Ms {\it Chandra}  observations in  the CDFS  were performed, we
here rely on the 4 Ms dataset  \citep[][ herafter X11]{xue},with a flux-limit  S$_{lim}
\sim$10$^{-17}$ erg s$^{-1}$ cm$^{-2}$ in the 0.5-2 keV 
(i.e. log(L)=42.6 erg/s @ z=6),  since it allows a more robust
comparison with published data. The additional observations in the CDFS are used as a posteriori test.

Throughout the paper we adopt a concordance $\Lambda$-CDM cosmology
 with $\Omega_{\Lambda}$=0.7, $\Omega_m$=0.3 and H$_0$=70 h$_{70}^{-1}$ km s$^{-1}$ Mpc$^{-1}$. Unless otherwise
 stated, errors are quoted at the 1$\sigma$ level. 
  
 \section{Observations and data analysis}
The 4Ms CDFS consists of 23  observations described in Table 1 of \citet{luo}
plus other 31 pointings described in X11 for a  total exposure of $\sim$4 Ms.
For the purpose of this paper we employed only observations taken with a focal temperature of 
$\le$-120 $^{\circ}$C since at higher T the background cannot be modeled
with our technique (see below). Differently from \citet{luo} and X11, because of higher detector temperature,
 we discarded {\em Chandra} OBS-ID 1431/0-1 ending up with a total exposure time of $\sim$3.8 Ms. \\
For every pointing, level 1  data  were reprocessed using  the {\em chandra\_repro} 
software in CIAO and   CALDB 4.6.1   released by the
{\em Chandra} team. Spurious signals from cosmic rays and instrumental features have been removed as well as
time intervals with flaring particle background. After cleaning, the effective exposure time is $\sim$3.6 Ms.
Astrometry has been improved by matching a high significance X-ray source catalog
with the Guo et al. (2013) catalog in the H  magnitude range 15$<$m$_{AB}<$23.
Images were created in the [0.5-2] and [2-7] keV energy bands, respectively.
In the same bands we created exposure maps at  effective energies of 1.2 and 3.2 keV, respectively.
Both images and exposure maps have a bin size of 0.5$\arcsec$. 
In the same energy bands we created background maps by using  the CXC
blank fields library. 
Above 9.5 keV the mirror effective area of {\em Chandra}  is basically zero; this means
that the  events accumulated at those energies are due to non cosmic (particle) interactions 
with the detector and the satellite. The level  of the non cosmic flux is variable because of several factors (e.g. Solar activity)
but its spectral shape  is  constant in time \citep{hm06}.  
Thus, in order to obtain a realistic particle background it is sufficient to rescale the maps in any band
by the ratio of the  [9.5-12] keV  number of events in the templates to the [9.5-12] keV number of events
in the real event file (see below for a more detailed treatment). 

While precise in estimating the particle background, this method may introduce 
a bias in the determination of the level of purely cosmic diffuse background.
Blank field event files contain a certain level of galactic background.
In fact, by construction 
blank field files are produced by averaging source-removed event files of extragalactic  fields
and randomizing  the position of remaining photons in order to remove background fluctuations 
clustering features \citep{cap12}. 
The  CDFS is a high  latitude field and its background is well approximated  in the blank field 
file library.  However since we assume that the particle background is well modeled by the
method above described, the level of galactic and solar system CXB
could be over- or underestimated. 
For that reason, after masking for X11 detected sources, we computed the following quantity
\begin{equation}
\Delta_{CXB}(E,d)=\sum_N(E,d)-\sum_N(E,b)
\end{equation}
where $\sum_{N}$(E,d) and $\sum_{N}$(E,b) are the total number of CXB photons in the energy band E
in the data and in the blank field files in any given pointing, respectively. 
This quantity, scaled to account for the source's masked area, is the number of over- or underestimated
local CXB photons in our maps. The $\Delta_{CXB}$ photons are  then redistributed across the field of view
and the detector according to the energy  dependent exposure map. In this way we expect a good
agreement between the real and the modeled background.
 A full description of the method can be   found in \citet{hm06}.
The images created with this method  suffer from Poisson random noise and cannot be 
adopted as background models. For these reasons the assembled mosaic of 
background maps  have been smoothed by using a Gaussian filter with $\sigma$=20$\arcsec$.

  \section{Source detection with {\em cmldetect} }
  \begin{figure*}[!t]
 \centering 
 \includegraphics[scale=.25]{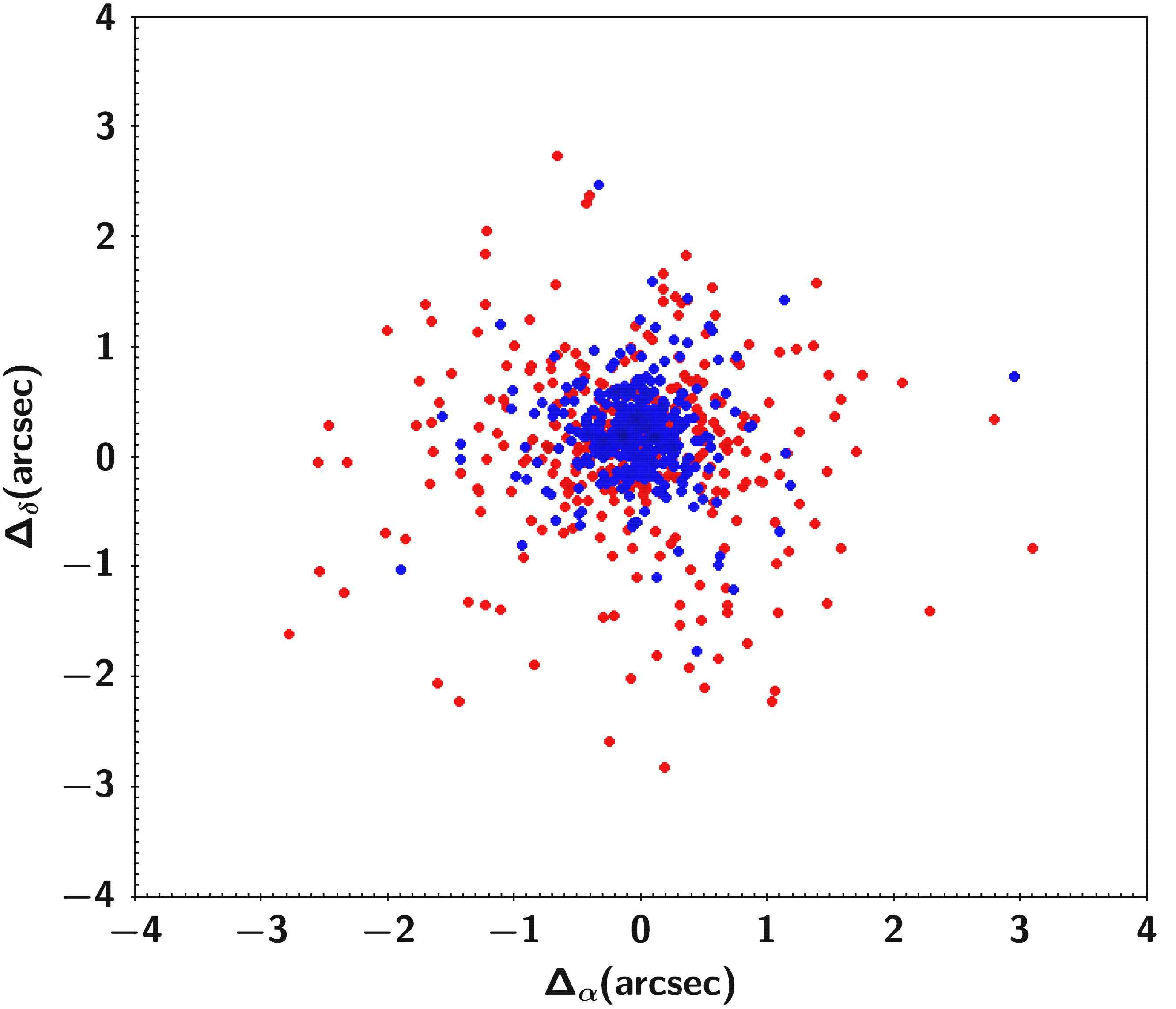}
  \includegraphics[scale=.4]{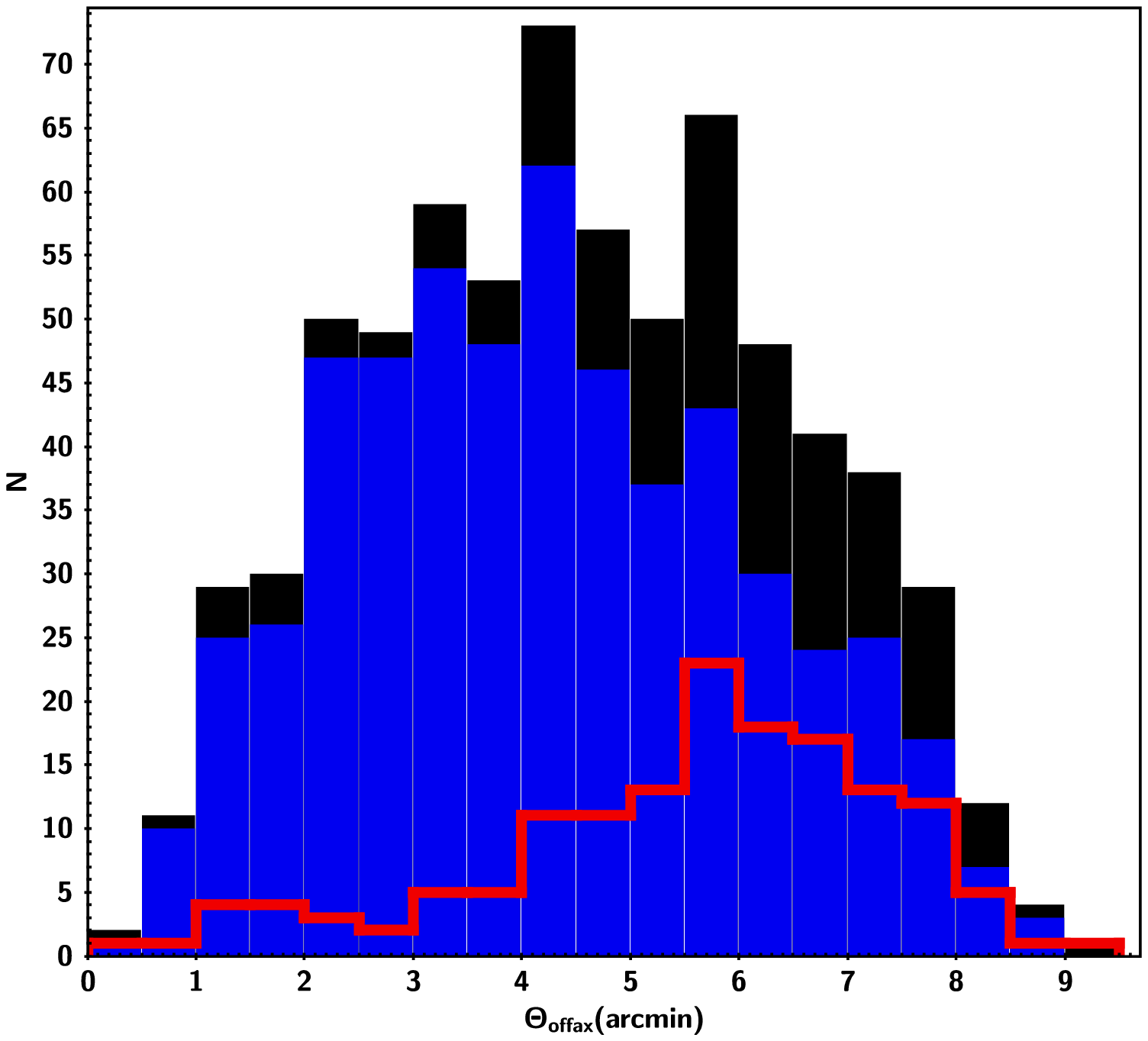}

\caption{\label{fig:shift} Left: The separation between the input position and the best fit X-ray centroid in arcsec In $red$ for sources with 
$\mathcal{L}<10$ and in $blue$ for those with $\mathcal{L}\geq10$. Right: the offaxis angle distribution
of the sources with an input vs output position smaller ($blue~filled~histogram$) and larger ($red~histogram$)  than 1$\arcsec$ compared
to that of the whole sample ($black~filled~histogram$).  }
 \end{figure*}

 Here we briefly summarize our source detection method and the main
 features of the detection software.  We employed a modified version
 of the XMM-SAS tool {\em emldetect}.  A description of the algorithm
 and of the statistical theory behind it can be found in \citet{crud}.
 While several authors have used {\em cmldetect} for analyzing {\em
   Chandra} surveys \citep[see e.g.,][]{pucc,kru}, the major step
 forward here is the employment of WFC3-HST galaxies as priors to
 improve the efficiency on faint sources and to facilitate the
 identification process.

This code  has been initially developed for ROSAT and XMM-Newton, and it was 
adapted \citep{pucc,kru}  for  use with {\em Chandra} with a customized version of the software {\em cmldetect} which makes
use of a {\em Chandra} PSF-Library and the XMM-SAS infrastructure.
Unlike the XMM-Newton PSF, the Chandra-PSF does not depend exclusively on energy and off-axis angle,
 but also on the azimuthal position. 
Such a feature cannot be handled by the XMM-SAS infrastructure; thus,
in order to allow the software to work with  {\em Chandra}, we created an $ad-hoc$ PSF library by averaging 
over all the azimuthal angles the PSF templates in energy and off-axis angle bins.
This approximation has been proved to be effective on several Monte Carlo simulations and on real data
within the {\em Chandra} COSMOS survey \citep{pucc}.  Moreover, since the geometry of the 4Ms CDFS mosaic 
is such that the roll angles are basically random, in this way the azimuthal PSF dependence is smeared out
and the approximation adopted in our PSF library carefully represents the real data. 

Given an input list of source positions,  simultaneous maximum likelihood PSF fits to the events 
distribution on the detector  are performed in all energy bands at the same time. Since the {\em Chandra}-CDFS 4 Ms
observations have aimpoints separated by $<$1$\arcmin$, we employ the cumulative mosaic
image and we fixed as a reference optical axis the mean  pointing position at $\alpha$=03$^h$ 32$^m$ 28$^s$.06, 
$\delta$=-27$^{\circ}$ 48$\arcmin$ 26$\arcsec$.

The most important fit parameters are: the source location, source extent  (beta model core radius), and source count rates.
Sources with overlapping PSFs are fitted simultaneously.  The maximum allowed number of sources 
that can be fitted simultaneously is  limited to 10, and it is
ruled by the  parameter   {\em nmaxfit} which sets the maximum number of sources
which are considered simultaneously. After some trial, we set 
{\em nmaxfit} =5 as a compromise between the deblending performances
and the computational times, that become impracticable for larger values.

Two parameters determine the image region on which a source fit is performed: 
{\em ecut} determines the size of the sub-image around each source 
used for fitting, and  {\em scut} determines the radius around each source,
in which other input sources are considered for multi-PSF fitting.
Both {\em ecut} and  {\em scut} are given as encircled energy fractions 
of the calibration PSF. For our purposes we fixed  {\em ecut}=0.68 {\em scut}=0.9 as in \citet{pucc}.
 
 All detection likelihoods are transformed to equivalent likelihoods $\mathcal{L}_2$ ($\mathcal{L}$) (see XMM $emldetect$ manual\footnote{\url http://www.cosmos.esa.int/web/xmm-newton/sas}), 
 corresponding to the case of two free parameters to allow comparison between detection runs with different numbers of free parameters:
\begin{displaymath}\mathcal{L}_2 = -\ln (1-P(\frac{\nu}{2},\mathcal{L}')) \;\;\;
{\rm with} \;\;\; \mathcal{L}' = \sum_{i=1}^n \mathcal{L}_i \end{displaymath}
where $P$ is the incomplete Gamma function, $n$ is the number of energy bands involved, $\nu$ is the 
number of degrees of freedom of the fit ($\nu =
3+n$ if task parameter fitextent=yes\footnote{If  fitextent=yes the sources are also 
fitted with  a convolution of beta or gaussian profiles with the PSF and if the likelihood obtained is significantly larger than that
obtained with the PSF only, the source is classified as extended}, and $\nu = 2+n$ otherwise), and $\mathcal{L}_i = C_i/2$ where $C$ is the statistics defined by \citet{cash}. 
The equivalent detection likelihoods obey the simple relationship  
\begin{equation}
\mathcal{L}_2 =- \ln(p),
\label{eq:eq1}
\end{equation} 
where $p$ is the probability for a random Poissonian fluctuation to have caused the observed source counts. 
 Note that for very small numbers of source counts (less than $\approx 9$ counts, Cash 1979), this relation likely does not hold and thus 
 the low count regime  must be tested with $ad-hoc$ simulations.

For this work, the input list for {\em cmldetect}   was made by
 the positions of the 34930  CANDELS GOOD-S WFC-3 selected sources \citep{guo13} on 
a total area of 0.048 deg$^{2}$.
The details of the parameters adopted and the properties of the
resulting catalogs are described later in Sect. 5. Here we focus on
the detection process and the association with the input priors.

As a first step, we
 fixed the source position  (parameter $fitposition$=no in {\it cmldetect}) to the input value, while the source flux was the only free parameter.  
 The fit was performed in the 
[0.5-2] keV and [2-7] keV energy bands simultaneously. Thus, by construction the equivalent likelihood from which we set the threshold
is that of  the [0.5-7] keV band. For our purposes we did not search for extended sources, thus we set $fitextent$=no.  
We first apply a preliminary threshold at $\mathcal{L}_{2}\geq$3 while the 
final threshold for the catalog is chosen only after the simulations (see below).  
Due to PSF blurring bright sources are observed on several pixels,
especially off-axis, the same X-ray source could be the counterparts of several 
CANDELS galaxies.
If there are more than 5 candidates with our Multi-PSF fitting software  
it could happen that at the location of
bright sources and on their PSF wings the software could find more detections. 
If the source is detected with more than 400 counts (i.e. $<$10\% of all 
the sources in the 4Ms CDFS, see below), 
within the 90\% of the PSF radius we keep only the detection with the
higher $\mathcal{L}$ and remove the other(s) from the catalog. 
At lower counts levels a visual inspection 
does not show any obvious case of multiple sources. 
 
Although the astrometry of {\em Chandra} is calibrated to be precise
within 0.5$\arcsec$, offsets between the X-ray and the near--IR
position may exist, and lead to additional errors in the determination
of the X-ray flux. To verify this effect and 
to provide the best possible coordinates for the X-ray centroid we
then released the constraints on the position of the X-ray emission
by letting 
 {\em cmldetect} run with $fitposition$=$yes$.
In doing so we realized that the internal structure of {\em cmldetect} software
loses track of the actual ID of the prior during the multi source fit
within the PSF encircled energy fraction parameters set by {\em scut}
and {\em ecut}.  Since this is a crucial information we had to correct for this effects $a-posteriori$ so, 
by inquiring the software developer\footnote{H. Brunner personal communication}
and after testing the procedure, we assigned  again the source
to the prior that is closer to the  X-ray centroid.  
This is not meant to assign a counterpart to the X-ray source, but 
simply to keep track of the input prior source.
However, we have also found that in some case
the revised position of the X-ray centroid is significantly shifted
with respect to the position of the original prior. This is
shown in Fig. \ref{fig:shift}, where we show the displacement between the best fit and input  CANDELS sources 
position.  We note
that for $\sim$80\% of the sources  the X-ray centroid is consistent 
with the position of the input source within 1$\arcsec$, although
there is however a tail at larger offsets (i.e. $\simeq 20\%$ at
$> 1.0\arcsec$ and $<10\%$ at $>$1.5$\arcsec$). 

This effect depends strongly on two quantities: the position on the
field and the X-ray intensity.
Indeed, as one can notice in the right panel of Fig. \ref{fig:shift},
the  majority of the sources with large offset are objects detected at low significance
($\mathcal{L}<10$) and at off-axis angles $>$4-5 $\arcmin$ (see Fig. 
\ref{fig:shift}).

This is not entirely surprising - it is well known that the image
quality of the Chandra images on the Goods--South field  degrades
significantly at large offset from the center, most notably due to a
significant degradation of the PSF, that leads to a lower positional
accuracy. It also indicates that the centering of X-ray sources
becomes difficult at low $S/N$. 


To investigate the origin of this shift we have visually inspected all the
relatively few ($\simeq 30$) sources that have an offset larger than
1'' but are also detected at good $S/N$ ( i.e. $\mathcal{L}>10$),
i.e. those for which the X ray position can be determined unambiguously. We have
verified that in most cases the large shift is due to some error in the
determination of the X-ray centroid, usually due to the poor PSF at
wide distances from the center (most of these sources are indeed close
to the image edges) or to tensions between the position in the soft-X
and hard-X images. In nearly all cases however the association with
the optical prior is robust, since the true X-ray center is actually close to
the optical center.  However, at this stage of the analysis, the association 
of a prior to a X-ray source should  not be  considered as an identification but 
simply as a test of the robustness of the procedure.

To better scrutiny the reliability of our procedure and the origin of
possible systematic effects we have designed a set of simulations and
a comparison with other approaches to source detection, that are
described in the following sections.

\section{CANDELS X-ray simulations} 
 The production of a source catalog requires
a deep knowledge of its statistical properties as well
as its limitations. 
In particular a fundamental property of a catalog is 
the selection function and the contamination from spurious detections. 
The best way to evaluate these characteristics is to  test the 
procedure on a sample of simulated source whose properties  are
known {\it a priori}. Also the instrument simulating carefully the property of
the instrument is  fundamental to evaluate the quality of the catalog.
In this section we present the statistical properties of our catalog
as well as validation of the quality of the method.
 \begin{figure}[!t]
 \centering 
 \includegraphics[scale=.4]{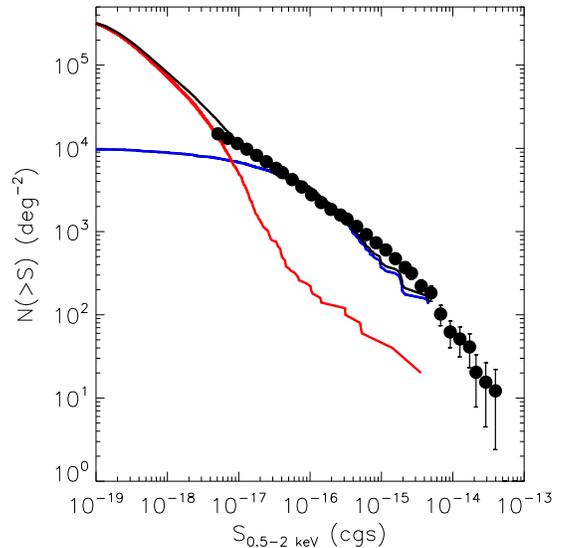}
\caption{\label{fig:logn} Comparison of the simulated [0.5-2] keV  cumulative number counts 
logN-logS for SFGs ($red~continuous~line$) and AGN ($blue~continuous~line$) 
with the measurements of \citet{leh} in the CDFS ($black~filled~circles$).
 The total model SFG+AGN is plotted as a $black~continuous~line$.} 
\end{figure}
\subsection{Simulated galaxies and AGN samples.}
Detecting X-ray sources using Optical/NIR priors is a relatively new 
procedure (see e.g. Fiore et al. 2012) which needs specifically designed simulations to 
validate its photometric accuracy and source detection yield. Every CANDELS galaxy
was assigned an X-ray flux and folded into a ray-tracing MARX  (Model of AXAF Response to X-rays) simulation 
to mimic the {\em Chandra} performances. 
In order to reproduce in a realistic way our mock sample
we created artificial X-ray fluxes of CANDELS galaxies 
from the estimated L$_{8-1000\mu m}$  by using {\em ad-hoc} scaling relations between
L$_{IR}$ and L$_X$ (see below). 
Infrared luminosities ($L_{\rm IR}$, from $8$ to $1000\,\mu{\rm m}$) are predicted for 
all galaxies in the catalog starting from their observed photometric redshift, their stellar mass \citep{santini},
 their $UVJ$ rest-frame colors and their observed (or extrapolated from the SED) 
   UV luminosity ($1500\AA$). We first split our sample into 
 actively star-forming and quiescent galaxies using the $UVJ$ color-color
  selection \citep{williams2009}. Quiescent galaxies are given zero $L_{\rm IR}$. 
  For star-forming galaxies, we predict their total SFR assuming that 
  they follow the redshift dependent SFR--$M_*$ correlation, the so-called ``main sequence'' of star-forming galaxies, 
  using the observed relation from \citet{score} and adding a 0.3 $dex$ random scatter, 
  mimicking the observed dispersion of the SFR--$M_*$ correlation.
   We convert the rest-frame UV luminosity into a 
   non-obscured SFR using the formula introduced in \cite{daddi2004}, and subtract
    it from the predicted SFR to recover only the dust-obscured component. Finally, 
    we convert this remaining SFR into $L_{\rm IR}$ using the formula of \cite{kennicutt1998}.
    In order to derive the X-ray luminosity of Galaxies we adopted the prescription of Basu-Zych (2013)  which 
  relates $z$ and     SFR to L$_X$ for star-forming-galaxies. Galaxies with a predicted 
 [0.5-2] keV flux $<10^{-20}$ (cgs) were flagged with S$_X$=0.

A fraction of CANDELS galaxies could be AGN which are powerful X--ray
sources. In order to include AGN X--ray emission in our sample,
we divided the sample in $\Delta$(z)=0.1 redshift bins, and in every bin we
 assigned an AGN flux (S$_{AGN}$) to a fraction of  galaxies
 consistent with that expected by the Gilli et al. (2007) population synthesis model down to
10$^{-20}$ erg s$^{-1}$ cm$^{-2}$. We point out that with this method
 the luminosity function of X-ray AGN is correctly reproduced, but the random choice
 of the AGN host galaxy does not allow us to obtain the correct optical/NIR luminosity 
 distribution of the simulated X-ray source counterparts. As a result,
 we may typically assign AGNs to galaxies that are fainter than the
 real AGN hosts.


 In Fig. \ref{fig:logn} we show the simulated logN-logS 
 of X-ray sources derived with this method compared with the 
 number counts measured by Lehmer et al. (2012). 
 
 \subsection{Ray-tracing events simulation \label{sect:ray}}

 To  simulate the CANDELS X-ray sources we employed the raytracing
 software MARX which provides a detailed ray-trace  simulation of  {\em Chandra} observations and can generate 
 standard FITS events files and images as output. It reproduces
 the {\em Chandra} mirror system and all focal plane detectors, including ACIS-I.
 The pointing direction, boresight, roll angle and dithering were reproduced to simulate all the 34930 CANDELS sources. 
 Every input source was assigned a photon X-ray spectrum modeled as 
 a simple power-law with $\Gamma$=1.4 plus Galactic absorption with N$_H$=7$\times$10$^{19}$ cm$^{-2}$ \citep{dl} and a normalization 
 derived  from its flux.  For every galaxy the software produces the expected number of events as a  function
 of energy by randomly drawing  them from their spectral distribution. Every photon has been spread on the 
 detector according to the actual PSF  template from calibration at any given energy and radial/azimuthal coordinates.  Detector response was
 reproduced within MARX, pixel randomization was also applied.    Dithering of the satellite  was also taken into account by using an internal MARX model. 
Since the software can handle one source and one pointing per run,
for every galaxy we produced 54 event files.  All the 34930 sources event files
 simulated over 54 pointings  were co-added and reprojected to the same tangent point.
For every pointing, the background in the energy band [E] has been estimated with the technique described by \citet{hm06} by randomly 
extracting events from the blank field background files  so that $B_{sim}[E]=\frac{B_{d}[9.5-12]}{B_{m}[9.5-12]} B_{m}[E]$ where, 
$B_{sim}[E]$, is the number of background events in the energy band [E], $B_{d}[9.5-12]$ is the number of events 
in the real data in the [9.5-12] keV energy band, $B_{m}[9.5-12]$ is the number of events in the blank field event files in 
the [9.5-12] keV energy band and $B_{m}[E]$ is the number of events in the  blank field event files in [E] energy bands, respectively. 
The sources and  the background simulations were then merged in a single event 
file and  images were produced. 

\subsection{Method reliability : source detection on simulated maps}
We use these simulations to  test the detection procedure and  to verify its efficiency.
Synthetic images were produced from the simulated  event files in the [0.5-2] keV, [2-7] keV and [0.5-7] keV 
energy bands with a spatial binning of 0.5$\arcsec$. 
Similarly we used the resampled blank field background maps described in Sect. \ref{sect:ray}
 to create background maps in the same energy band and with the same
spatial binning as in images. Background maps were smoothed with a Gaussian kernel with $\sigma$=20$\arcsec$ 
As exposure maps we employed those computed for the real data.  

We ran a source detection on the simulated images with  the same parameters of the 
real data.  In the real data, in $\sim$20\% of the cases, the actual
detected source is found more than 1$\arcsec$ away from the galaxy flagged as 
prior. By making use of our simulations we  checked 
this fraction and found the same result. 
We first notice that the values of $\mathcal{L}$ of most of the detected sources improves significantly
by  fitting of the position (i.e. by using $fitposition=yes$ compared to $fitposition$=no). 
As in the real data, the fraction of sources for which  we find a $>$1$\arcsec$ displacement
 between the prior and the best fit X-ray centroid shows a strong 
radial dependency. At offaxis angles $<$4-5$\arcmin$, the number of such sources
is of the order of 5\% while, at larger offaxis angles, this fraction is of the order of 30\%.
Since the only difference between center and off-center in the
simulations is the degraded PSF, we conclude that a larger fraction 
of the X-ray centroids  at relatively large off--axis angles are significantly displaced from their prior
due to the PSF degradation.

We can use the simulations to verify the accuracy of our procedure in
determining the correct prior. This is not straightforward since in
our simulations a X-ray flux is assigned to all the star--forming
galaxies in the input sample. Most of them have fluxes very small,
definitely below the detection limit, but also non-zero. To take this
into account we used the statistical approach used in \citep{cap07},
that compares the input and output catalogs of the simulations using
the match in both position and flux. We evaluated how many "prior" sources are the actual 
counterpart of the  detected X--ray sources by cross-correlating our
output catalog with the  input one by  minimizing the following quantity \citep{cap07}:
 \begin{equation}
 R^2=\left(\frac{X_{out}-X_{in}}{\sigma_{X,out}}\right)^2+\left(\frac{Y_{out}-Y_{in}}{\sigma_{Y,out}}\right)^2+\left(\frac{S_{out}-S_{in}}{\sigma_{S,out}}\right)^2
 \end{equation}
where X,Y are the coordinates on the detector and S is the flux in the [0.5-7] keV band, respectively. 
This estimator is also known as Mahalanobis distance \citep{book}.
The subscripts  $in$ and $out$ stand for input and output catalogs, respectively.
As a first result, we find that  for $\sim$2\% and $\sim$8\% of the
detected sources on-- and off--axis respectively,   the actual
counterpart is not the prior.

\begin{figure}[!t]
\center
\includegraphics[scale=.7]{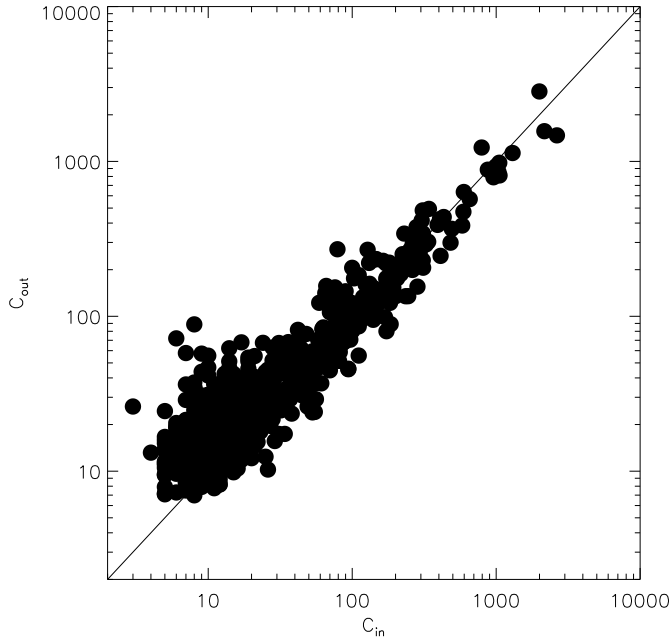}
\caption{\label{fig:counts}  Photometry efficiency test on the simulations. The input versus output source counts.   }
\end{figure}

We also tested the accuracy of the photometry: in Fig. \ref{fig:counts}  we show the [0.5-2] keV input vs output counts. 
As in \citet{pucc}  the output/input counts ratio is consistent with 1
and spread according to a Poisson distribution. At faint fluxes the
distribution appears to be 
skewed toward high C$_{out}$/C$_{in}$ ratios because of a sort of
Malmqvist  bias'' - i.e. we do not plot in Fig. \ref{fig:counts}
objects with a low $\mathcal{L}$ parameter.

\begin{figure}[!t]
 \center
 \includegraphics[scale=0.7]{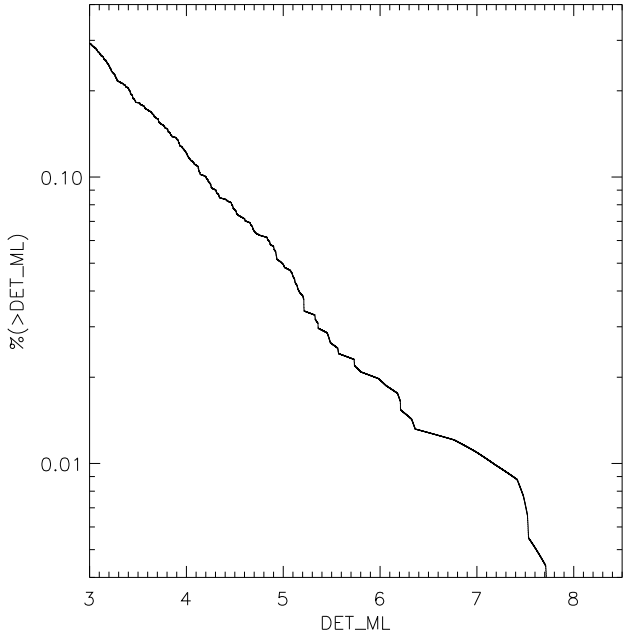}
\caption{\label{fig:detml} The fraction of spurious detections in the GOODS-S field as a function of the
detection likelihood as determined by our Monte Carlo simulations in the [0.5-7] keV band.  } 
\end{figure}

These simulations are able to guide us in the choice of a crucial
parameter, namely the detection threshold.
At this  aim, we have to compute the expected
number of background fluctuations detected as sources as a function of the detection likelihood $\mathcal{L}$.   
We did this by running a source detection using as X-ray map a randomized image of the modeled background and the CANDELS catalog 
as input. In this way   the number of detections can be considered an
estimate of the overall number of spurious detections in the real data. In Fig. \ref{fig:detml}  we show the cumulative 
 distribution of the ratio between the spurious sources detected in
 these simulations and the real sources detected in the data as a function of the $\mathcal{L}$ parameter.
Since the goal of this paper is to push the limit of deep fields
beyond the actual one and maximize the detection of faint sources, we
estimate that an acceptable spurious fraction should not be higher
than 5\%, compared to the usually adopted values of $\sim$1--2\%.   
This fraction corresponds to values $\mathcal{L}$$>$4.98 and
translates into a minimum flux detection significance of
$\sim$2.7$\sigma$ (Eq. \ref{eq:eq1}). This  is similar to the 
value reached with blind detections at comparable background levels
\citep{luo,xue}.

Finally, we checked if the simulated background carefully represents
the actual level. In fact, we know  that the real background 
fluctuations \citep{cap12} are not randomly distributed, but  are
strongly correlated. 
On the other hand the simulated background is relatively smooth and
uniform and this could introduce a bias in the spurious fraction estimate.
For that reason we performed a  source detection on the real data 
masked for the detected  sources according to the PSF size at the source location. 
The umasked part of the image can be considered as a fair
estimate of the real background. We have then produced a catalog of 34930  positions 
drawn from the real catalog by randomly placing the artificial sources in an annulus with inner
 and outer radii 5$\arcsec$-10$\arcsec$ from the real prior sample of
sources, respectively. In this way we preserve the spatial  distribution
 of the CANDELS sources in the input catalog but we do not
overlap with real sources.    We then ran our source detection on this masked image
by using as input catalog the random sample above described. We repeated such
a procedure 20 times. All these  detections are nothing else but random background fluctuations 
which would enter the catalog as spurious sources.
The results found with this test are fully consistent 
 with those obtained with the randomized background images.
  
\begin{figure}[!t]
 \center
  \includegraphics[scale=.7]{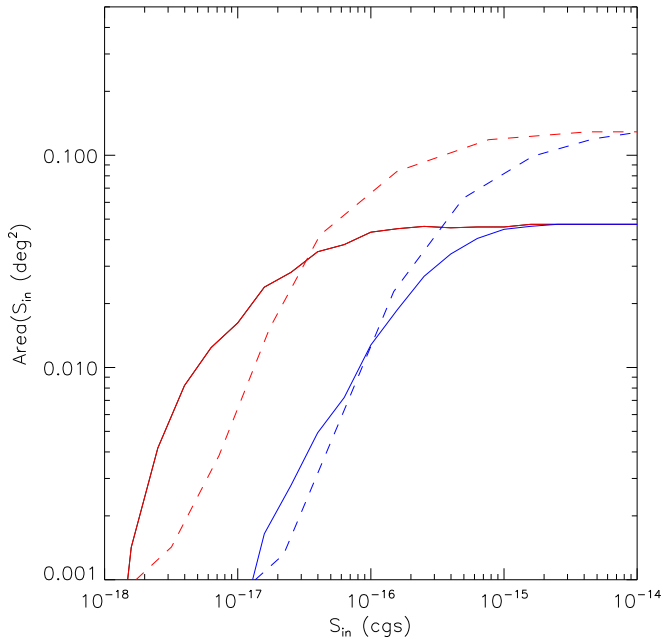}
\caption{\label{fig:scov} The sky area vs input flux selection function plot for our sample in two sub-bands compared with that of
\citet{leh}. The   $red-continuous$ and the $blue-continuous$ line represent the selection functions in the  [0.5-2] keV and [2-7] keV energy bands, respectively. The
$red-dashed$ and the $blue-dashed$ line represent the selection functions  from \citet{leh} in the  [0.5-2] keV and [2-7] keV energy bands, respectively. Hard band fluxes 
have been extrapolated to [2-10] keV fluxes }
\end{figure}

We then computed the selection function of our detection procedure
 by evaluating the ratio of
the number of retrieved input  sources with respect to that of 
input ones in bins of intrinsic input flux of 
$\Delta\log(S_{in})$=0.1. The resulting cumulative histogram is smoothed 
 with a filter width of $\delta\log$S=0.3. The final sky coverage is  shown in Fig. \ref{fig:scov}.
Note that here we present the sky coverage with respect to the intrinsic (and not the detected) flux of
the X-ray sources. 
 
The results are compared with those  of \citet{leh} obtained with
a Bayesian method for flux calculation and for blind X--ray source
detection in the CDFS. 
As expected, the faintest recovered sources detected with the two methods have a similar flux, 
but our method yields a steeper selection function at faint fluxes. As an example, in the [0.5-2] keV band,  with a thresholds $\mathcal{L}>$4.98 (see below) in the faintest
fractions of decade of fluxes  our method can recover about a  factor 5 more sources. This is particularly evident in [0.5-2] keV energy band, but not as much in the [2-7] keV band. This
is due to the fact this method take advantage of highest angular resolution of {\em Chandra} at low energies. 

In summary, in this work we have explored the advantages of using a
prior-based search for X-ray sources in the GOODS-South field, isuing
teh cmldetedct software. 
These evidences allow us to draw the first conclusions about the quality of this method: 
a) at offaxis  angles $<$4$\arcmin$ for  98\% of the sources the prior galaxy is likely to be the counterpart to the X-ray source.
b) at offaxis  angles $>$4$\arcmin$ (i.e. if the PSF HEW$>$1.5$\arcsec$) the prior sources and the relative detected X-ray sources are 
significantly displaced in 20\% of the cases, but  for 92\% of the sources the prior galaxy is likely to be the counterpart to the X-ray source.
c) the source detection quality is improved by fitting in any case the position of the X-ray centroid, 
c) using  a deep optical catalog as a prior, increases the probability
to detect a faint X-ray source compared to that of a blind detection based on background fluctuations.
To some extent, the limitations in this approach are certainly due to
the complex nature of the X-ray data in the CDFS area, that degrade at
large distances from the centre. However, some of these limitations
can be due to the specific performances of cmldetect, that was not
originally designed to be used in this way. In future works we plan to
adapt other prior-based software for photometry (like T--PHOT, Merlin
et al 2015)  to the case of X--ray data.

\section{X-ray  Catalog Assembly}

Armed with the results  of the simulations descriobed above, we have
obtained the final catalog in the GOODS-Sout field. We summarize in
this section the procedure finally adopted and the comparison with
other approaches.

\subsection{The prior-based catalog}
We run the source detection on the 4 Ms CDFS data  [0.5-2] keV and [2-7] keV band simultaneously and
 the likelihood is computed in the  [0.5-7] keV band. We used as input catalog the positions of the
34930 sources detected by \citet{guo13} in the CANDELS GOODS-S area and set $fitposition$=no and 
we imposed a $\mathcal{L}$=4.98 threshold in the resulting [0.5-7] keV energy band. 
In this way we preselected 735 sources, some of which corresponding to the same X-ray source. 
 We then fitted the position of the sources
to determine the best possible X-ray centroid of each detected source. 
At this threshold we detect 698 unique sources, in the $\sim$0.048 deg$^{2}$ of the CANDELS GOOD-S area
analyzed  by \citet{guo13}. We considered only point sources and we did not fit the extension of the sources.
Source falling within  the region of groups/clusters detected by \citet{fin} were visually inspected individually.
For every source, we determine the source counts and the count-rate as an output 
of the detection algorithm, the background level, the PSF 90\% Encircled Energy Fraction (EEF) and the 
$\mathcal{L}$ in the  [0.5-2] keV, [2-7] keV  and [0.5-7] keV energy bands, respectively. 
Count rates were converted into fluxes by 
assuming a simple power-law  spectrum with $\Gamma$=1.4 plus a Galactic absorption   $N_{\rm H} =7 \times 10 ^{19}$ cm$^{-2}$ \citep{dl}. 
The Energy Conversion Factors (ECFs) were computed with the online tool {\em Chandra} PIMMS.  
 The response of the ACIS-I detector varied significantly across the {\em Chandra} lifetime, for this reason we
computed the ECFs for every pointing's epoch and then weight-averaged them according to the exposure time. 
As a result we obtained a count-rate to flux ECF of 5.32$\times$10$^{-12}$ erg cm$^{-2}$ in the [0.5-2] keV band and 
2.71$\times$10$^{-12}$ erg cm$^{-2}$ to convert the [2-7] keV count-rate into  a [2-10] keV  flux.
The full band count-rate, counts and fluxes are the sum of those in the two sub-bands, respectively.
As mentioned above  the overall significance of the detection is measured with   the cumulative  [0.5-7] keV energy band net counts, thus for 
some sources the parameters in the [0.5-2] keV, [2-7] keV  sub-bands may not be accurate.  For this reason the flux of the sources 
for which the sub-band detection has a significance lower than the threshold ($\mathcal{L}$$<$4.98), in the specific sub-band 
should be used with care.
 While all the sources have $\mathcal{L}\geqslant$4.98 in the [0.5-7] keV band, we report 534 and 285 significant 
 detections  in the  [0.5-2] keV and  [2-7] keV energy bands, respectively. We define these sources as N($\mathcal{L}\geqslant$4.98) in Tab. 1.
Among these 352 sources are detected in the [0.5-2] keV but not in the  [2-7] keV  band, 106 sources in the [2-7] keV but not in the  [0.5-2] keV  and only 61 sources
 have a significant detection in the [0.5-7] keV energy band and no significant counterpart in the sub-bands N($\mathcal{L}\geqslant$4.98).
 In Table \ref{tab:ns} we briefly summarize the properties of the X-ray catalog presented here. 
 \begin{table}
\center
\caption{Number of detections\label{tab:ns}}
\tablenum{1}

\begin{tabular}{l r r r }
\hline
 \multicolumn{1}{c}{      } &  
  \multicolumn{1}{c}{[0.5-2] keV } &
  \multicolumn{1}{c}{[2-7] keV} &
  \multicolumn{1}{c}{[0.5-7] keV} \\
  \hline
N($\mathcal{L}\geqslant$4.98)   &  531& 285 &  698\\
n($\mathcal{L}\geqslant$4.98)   &  352& 106 &  61\\
N(X11)& 466  & 254 &  527\\
N(X11+C15)& * &  * &  784\\
S$_{lim}$ & 0.11 &    0.87 & 0.89 \\
 \hline\end{tabular}\\
\tablecomments{ \label{tab:ns} From top to bottom: $N(\mathcal{L}\geqslant4.98)$  is the actual number of significant detections
in the three  energy bands; $n(\mathcal{L}\geqslant4.98)$  is the  number of sources significantly detected in a given energy band only (plus full band);
$N(X11)$ is the number of X11  significant detections
in the three  energy bands;
$N(X11+C15)$ is the total number of unique X-ray sources detected in the CANDELS GOODS-S area 
by X11 and in this work; S$_{lim}$  is the flux limit in each band in units of $\times$10$^{-16}$ erg cm$^{-2}$ s$^{-1}$.
}
\end{table}

\subsection{The comparison with previous catalogs}
In the same area X11  detected 527 X-ray sources by using the same X-ray dataset. 
They used a purely blind X-ray detection  without prior knowledge of the actual counterparts. 
Among these 466, 254 and 527 are detected in the [0.5-2] keV,  [2-7] keV 
and [0.5-7] keV band, respectively (N(X11) in Tab. 1).  A simple positional match between the two catalogs of their catalog
with a  2$\arcsec$ matching radius, returns  
443 sources in common, 252 detected with our method only and 85 detected only by X11.
In Fig. \ref{fig:sep} we show the distribution of the distances between the X-ray centroids found here and those of X11:
the average shift is $\sim$0.5$\arcsec$.
By merging our catalog with that of X11 we bring the total number of
X-ray detected sources in the CANDELS-GOOD-S area to 784.

 \begin{figure}[!t]
 \center
  \includegraphics[scale=.5]{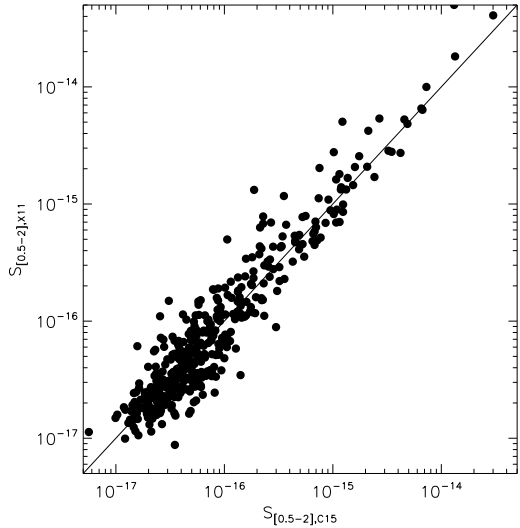}
  \caption{\label{fig:compxue} The [0.5-2] keV fluxes measured here compared with those of X11 for the common, above threshold sources.} 
  \end{figure}
 As a safety check we cross matched the counterpart catalog of \citet{hsu} with 
ours  for the 443   sources in common with X11.
 If we consider all the sources with a secure association in our
catalog, we find the  same association in 90\% of the cases. Three quarter of the 
remaining  have an  offaxis angle $>$4$\arcmin$. The likely reason of this discrepancy can 
 be the different method used for the  X-ray centroid estimate  with our method and the
  completely different method adapted by \citet{hsu} for assigning the counterpart to the 
X-ray sources. 
We compared  the fluxes properties of the 443 sources in common with those presented by X11. 
In Fig. \ref{fig:compxue} we show the comparison of the [0.5-2] keV fluxes measured by us and those of   X11.
There is a very good agreement between the measurements and the mean of the ratio of the
two measurements is $\sim$0.98. 
Our count-rate to flux convertion  (that uses a fixed spectral slope)
is different from that of X11,  who use for each source
a spectral index  obtained from the hardness ratio. This leads to an
intrinsic dispersion in the two measurements that 
has no a clear trend with flux. 

We also checked the 85 sources detected by X11 only. Among them 62 have been detected by 
our software,  but with 3.00$<\mathcal{L}<$4.98 and thus did not satisfy the selection criterion for being included in the catalog.
The remaining 28 unmatched sources are all at the very faint limit of
their catalog. Therefore  28/571 X11 sources are not found with our method even at $\mathcal{L}>$3.
We can explain this small fraction of ``missed'' sources with statistical fluctuations 
among the two catalogs or, alternatively they could belong to the
sample of extended  sources \citep[see e.g.][]{fin}. 

We also  performed a visual inspection of the
newly detected sources in this paper with the public deeper
observations in the CDFS.  At the time of writing $\sim$5.9 Ms of
data are available  in the archive. Among the 698 sources detected in
this work  only a handful of very faint objects seem to be undetected 
by visual inspection. Their number is consistent with the expected
spurious fraction (5\%).

\subsection{Validation of the prior matching}
 \begin{figure}[!t]
 \center
  \includegraphics[scale=.5]{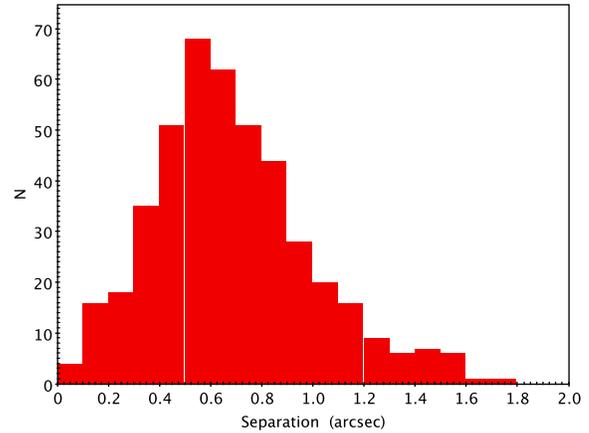}
\caption{\label{fig:sep} Angular separation between the 443  sources in common  with the catalog of X11}
\end{figure}
As we have shown above, our method potentially suffer from some
uncertainties, as shown by the relatively large fraction of objects
that are detected at large distances from the priors,especially for
faint sources at off-axis angle$>$4$\arcmin$. 
According to our simulations a fraction of the detected
X-ray sources may not be associated to the input prior at large
off--axis angles.

It is therefore interesting to explore the more traditional technique
for identifying counterparts of X-ray sources without priors, namely the  likelihood ratio
technique of \citet{suth}.
We followed the procedure of \citet{bru05,bru07} adapted for {\em Chandra} by \citet{civ10}.   For a
given  candidate counterpart with magnitude $m$ at a distance $r$ from the X-ray source, 
the likelihood ratio LR is defined as the ratio between the probability that the source is the correct identification and 
the corresponding probability for a background, unrelated object 
$LR=\frac{q(m)\,f(r)}{n(m)}$, where q(m) is the expected magnitude m distribution function  of the real optical
 counterpart candidates, f(r) is 
 a two-dimensional Gaussian probability distribution function of the positional errors, and n(m) is the 
 surface density of background objects with magnitude m.
The distribution of the local background objects, $n(m)$, was computed from each of the three input catalogs using the objects 
within a 5$\arcsec$--10$\arcsec$ annulus around each X-ray source. We chose a 5$\arcsec$ inner radius in order 
to avoid the presence of true counterparts in the background distribution, and a 10$\arcsec$ outer radius to exclude
 the counterparts of other nearby X-ray sources.  

The function q(m) has been estimated from our data as follows.
 We first computed q'(m) = [number of sources
 with magnitude m within 3$\arcsec$] - [expected number of background sources with magnitude m in a 3$\arcsec$ circle]. 
The choice of a 3$\arcsec$ radius is dictated by the requirement of maximizing the statistical significance of the overdensity around
 the X-ray sources. A smaller radius would include in the analysis only a fraction of the true identifications and the q(m) distribution 
 would be more affected by Poissonian noise. A larger radius would increase the number of background sources.
 
\begin{figure}[!t]
\center
\includegraphics[scale=.60]{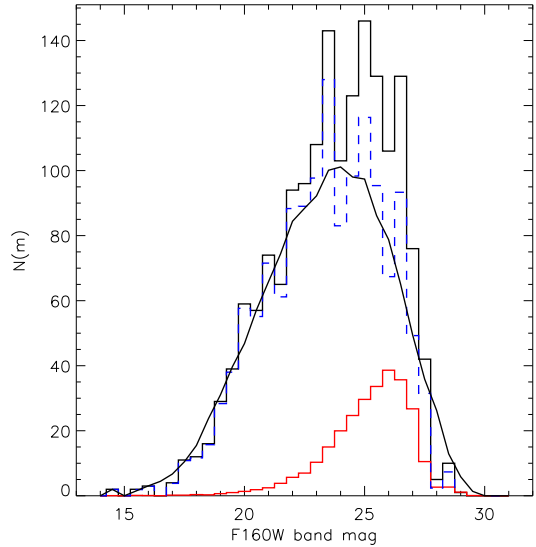}
\caption{\label{mag}  $Top:$ $Black~solid~histogram$ the magnitude distribution of all the \citet{guo13} sources within 3$\arcsec$  
from our X-ray centroids. $Red~solid~histogram:$ the expected background magnitude distribution of sources in an annulus with 
inner radius of 5$\arcsec$ and outer radius of 10$\arcsec$ from the X-ray source. The $blue~dashed~histogram$ is the resulting, non normalized, q(m) distribution  adopted
to compute the LR.  The $Black~continuos$ line is the adopted $q(m)$. }
\end{figure}

 As  extensively described in \citet{bru07}, with this procedure $q(m)$ is underestimated at faint magnitudes. 
 At fainter magnitudes, the number density of
 CANDELS sources  within the 
 search radius of each X-ray source is artificially smaller than that expected 
 from the whole sample $n(m)$. The reason for this biased estimate is the presence of a large number of moderately bright CANDELS 
 counterparts  within the X-ray centroids. These sources could occupy a non-neglible fraction  of the X-ray counterpart search area,  making 
 difficult to detect faint background objects. 
 Such a bias would produce an unrealistic negative $q(m)$, which 
 would prevent us from using the LR procedure at faint magnitudes. 
  In order to correctly estimate  $n(m)$ at faint magnitudes, we have randomly extracted from the CANDELS catalog 1500 NIR sources with the same 
 expected magnitude distribution of the X-ray source counterparts. Then we computed the background surface density around these random sample of galaxies.
Indeed, we found that  the $n(m)$ computed in this way is consistent with the first measured $n(m)$ at F160W$<$24.5 and much smaller than it 
 at faint magnitudes. 
 Therefore, the input $n(m)$ in the likelihood procedure was the global one for F160W$<$24.5   and that derived with this analysis  for F160W$>$24.5. 
 This allowed us to associate several very faint counterparts to X-ray sources that would have been missed without this adjustment to the procedure.
In Figure \ref{mag}, we show the observed magnitude distribution of the objects in the  1.6 $\mu$m catalog  within a radius of 3$\arcsec$ 
around each X-ray source (solid histogram), plotted together with the expected distributions of background objects 
in the same area ( red solid histogram). The smoothed difference between these two distributions 
is the expected distribution of the counterparts ($q'(m)$, black curve) before normalization. The $q(m)$ is obtained by normalizing $q'(m)$ to 1.  

For the probability distribution of positional errors, f(r), we adopted a Gaussian distribution with standard deviation, 
$\sigma=\sqrt{\sigma_{opt}^2+\sigma_X^2}$, where $\sigma_{opt}$ is the positional uncertainty 
for the optical sources that we assumed to be 0.1$\arcsec$ for all the sources. 
$\sigma_{X}$ was set to $RADEC\_ERR $ which is the error in the centroid provided by {\em cmldetect}.
 The $RADEC\_ERR $  in our catalog spans from  $\sim$0.1$\arcsec$ to $\sim$1.5$\arcsec$
We also added a 0.25$\arcsec$ systematic (half {\em Chandra} pixel) to take into account pixelation effects.
Having determined the values of q(m), f(r), and n(m), we computed the LR value for all the sources within 3$\arcsec$ of the 698  X-ray centroids.
As in \citet{civ10} and \citet{bru05} we had  to choose the best likelihood threshold value (L$_{th}$) for LR to discriminate between spurious and real identifications. 
L$_{th}$ must be small enough to avoid missing too many real
identifications, so that the completeness of the sample is high and
large enough to keep the number of spurious identifications low and
increase the identification reliability.
Extensive simulations indicate that the trade--off is obtained for
$R=C\sim$0.89 corresponding to  $L_{th}$=0.75.
As a result 698 sources have at least a counterpart within the search radius, but only for 608 the association  passes the LR test. 
With this threshold 529 X-ray sources have 1 significant counterpart with LR $>L_{th}$, 74 have 2 and
9 have 3 counterparts, respectively. For 90 sources we do
 not have a significant counterpart association and they are flagged with FLAG\_ASSOC=2 in the catalog.
\begin{figure}[!t]
\center
\includegraphics[scale=.70]{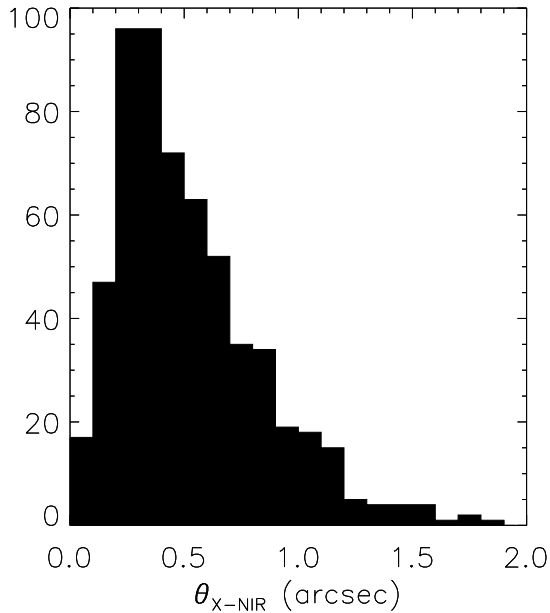}
\caption{ \label{fig:sep} The distribution in arcsec of the distances between the X-ray centroid and the  optical 
counterpart for secure identifications. }
\end{figure}
However,  in many cases, having multiple counterparts does not imply
that the identification is unsecure.  In order to resolve multiple
associations, we computed the distribution of the LR  among the
possible counterparts of the same X-ray source \citep{civ10}.

If such a ratio is larger that the median (LR$_{max}$/LR$_{i}$)
then we define the association as secure. In other cases the association is flagged  (FLAG\_ASSOC=-1) as ambiguous and the CANDELS ID number of
all the candidate counterparts  is listed in the catalog in LR order. Secure identifications are flagged with FLAG\_ASSOC=1.
After this procedure we have 552 secure identifications 57 ambiguous (double to triple) and 
89 are unsecure identifications and 3 unidentified (likely spurious X-ray detections). In Table \ref{tab:id} we summarize
the results of our identification procedure. As expected, we observe that the fraction of ambiguous and unsecure
identifications increases with the offaxis angle.
 \begin{table}[!h]
\center
\tablenum{2}
\caption{ The results of our LR identification procedure. \label{tab:id} }
\begin{tabular}{l r r  }
\hline
 \multicolumn{1}{c}{  Class    } &  
  \multicolumn{1}{c}{Number} &
  \multicolumn{1}{c}{ \%} \\
  \hline
  Secure & 552 & 79.1\% \\
  Ambiguous &  56 & 8.0\%\\
  Unsecure & 90 & 12.9\%\\
   \hline\end{tabular}\\
\end{table}  

In Fig. \ref{fig:sep} we show the distribution of the distance between 
the X-ray centroid and the best counterpart in the CANDELS catalog, 
this distribution peaks at $\sim$0.25$\arcsec$ and sharply declines
down to 2$\arcsec$. 

We can finally compare the results of the prior-based photometry with
this likelihood ratio technique. 
We find that the results are nicely consistent. Indeed,  545/552
($\sim$98.7\%) of the secure identifications are associated with the
input prior CANDELS ID and 43/57 in the case of ambiguous sources.
 We note that for this comparison cannot be performed over 90/698
 sources, i.e. 13\% of the sources, for which the LR does not yield
 any result.
 
We point out that the majority of the sources for which the
counterpart is flagged as unsecure and is not coincident with  prior,
are found, on average, with $\mathcal{L}<$10 and at large off-axis angles and thus with a broad  PSF.
In particular at off-axis angles $<$4$\arcmin$ the fraction of sources for which the counterpart is not the prior 
is $<$1\% while at off-axis angles $>$4$\arcmin$ this is $\sim$9\%.
We added in the catalog  a flag, FLAG\_PRIOR which has value 1 for off-axis angles $<$4$\arcmin$ or  $\mathcal{L}>$10
and 2 for off-axis angles $>$4$\arcmin$ and  $\mathcal{L}<$10. If FLAG\_PRIOR=1 one can 
safely use the prior source as the actual counterparts. Otherwise, one should check if the results of the 
LR yields to another counterpart.
  In our simulations the AGN X-ray flux is randomly assigned to a CANDELS Galaxy thus  we could not
test the LR because the AGN magnitude distribution was the same of that of background sources. 
If a source with no prior was simulated it would not be detected however, the only source of 
potential errors is the high probability  that a source is detected by chance given a random prior within $ecut$. 
To evaluate this we performed the following test: 
to avoid contamination by bright sources we selected 1847  prior 
candidates within 4 $\arcsec$ of the 500 faintest detected sources. From that catalog we removed
the sources which we identified as "BEST\_ID" and run the source detection on 455
of them who have more than one counterpart. 
We removed a posteriori from the 1847 input sources the actual counterpart of each X-ray source
and run the source detection.
As a result we have detected only 169/455 detection above threshold
with $\mathcal{L}>$4.98. For these sources the recovered X-ray centroid is consistent with 
that obtained with the master prior catalog. We repeated the LR test and
for  99.5\% of the secure matches the best candidate was still BEST\_ID. While 
an evaluation of $ecut$ is not straightforward, we notice that the sources not
detected by this test are, as expected, those whose prior had a distance from the X-ray centroid
larger or similar to $ecut$.
\subsection{Catalog columns description} 
Our catalog is available  in machine readable format at the URL
\url{http://www.astrodeep.eu/data/} and on Vizie-r. 
Here we describe  the   
columns in the online catalog.\\
NID: ID of the X-ray source. \\
PRIOR\_ID: CANDELS ID of the  optical source used as prior for the X-ray source detection.\\
FLAG\_PRIOR: Flag to determine the reliability of the association with a prior.\\
BEST\_ID:CANDELS ID of the primary optical counterpart of the X-ray source from LR.\\
SECOND:CANDELS ID of the second best optical counterpart of the X-ray source from LR.\\
FLAG\_ASSOC: Quality of the identification flag.\\
RA\_X: Best fit right ascension in decimal Degree units of the X-ray centroid.\\
DEC\_X: Declination in decimal Degree units  of the X-ray centroid.\\
RADEC\_ERR: 1-D error on the  X-ray centroid position (arcsec).\\
SEP: Distance from the best optical counterpart\\
SCTS\_FULL: [0.5-7] keV counts.\\
SCTS\_FULL\_ERR: 1 $\sigma$ [0.5-7] keV counts error.\\
SCTS\_SOFT: [0.5-2] keV counts.\\
SCTS\_SOFT\_ERR: 1 $\sigma$ [0.5-2] keV counts error.\\
SCTS\_HARD: [2-7] keV counts.\\
SCTS\_HARD\_ERR: 1 $\sigma$ [2-7] keV counts error.\\
$\mathcal{L}$\_FULL: -ln(p) determined in the [0.5-7] keV band.\\
$\mathcal{L}$\_SOFT: -ln(p) determined in the [0.5-2] keV band.\\
$\mathcal{L}$\_HARD: -ln(p) determined in the [2-7] keV band.\\
FLUX\_FULL:[0.5-10] keV flux in erg cm$^{-2}$ s$^{-1}$ in units 10$^{-16}$.\\
FLUX\_FULL\_ERR 1$\sigma$ :[0.5-10] keV flux error in erg cm$^{-2}$ s$^{-1}$ in units 10$^{-16}$.\\
FLUX\_SOFT:[0.5-2] keV flux in erg cm$^{-2}$ s$^{-1}$ in units 10$^{-16}$.\\
FLUX\_SOFT\_ERR 1$\sigma$ :[0.5-2] keV flux error in erg cm$^{-2}$ s$^{-1}$ in units 10$^{-16}$.\\
FLUX\_HARD: [2-10] keV flux in erg cm$^{-2}$ s$^{-1}$ in units 10$^{-16}$.\\
FLUX\_HARD\_ERR 1$\sigma$ :[2-10] keV flux error in erg cm$^{-2}$ s$^{-1}$ in units 10$^{-16}$.\\
RATE\_FULL:[0.5-7] keV count rate  in ph cm$^{-2}$ s$^{-1}$.\\
RATE\_FULL\_ERR 1$\sigma$ :[0.5-7] keV count rate error in ph cm$^{-2}$ s$^{-1}$.\\
RATE\_SOFT:[0.5-2] keV count rate  in ph cm$^{-2}$ s$^{-1}$.\\
RATE\_SOFT\_ERR 1$\sigma$ :[0.5-2] keV count rate error in ph cm$^{-2}$ s$^{-1}$.\\
RATE\_HARD:[2-7] keV count rate  in ph cm$^{-2}$ s$^{-1}$.\\
RATE\_HARD\_ERR 1$\sigma$ :[2-7] keV count rate error in ph cm$^{-2}$ s$^{-1}$.\\
HR1: Hardness ratio.\\
HR1\_ERR: Hardness ratio error.\\
OFFAX: Off Axis Angle in arcmin.\\
RA\_OPT: Best fit right ascension in decimal Degree units of the best CANDELS counterpart.\\
DEC\_OPT: Declination in decimal Degree units  of the best CANDELS counterpart.\\
m160: F160W AB magnitude.\\
Spec\_z: Spectroscopic redshift from \citet{santini}. \\
Photo\_z: Photometric redshift from \citet{santini}.\\
Photo\_z\_H: Photometric redshift from \citet{hsu}.\\
X11: Source ID in X11\footnote{sources in the X11 supplementary catalog have been number with their ID+1000}.\\
H14: Source ID in \citet{hsu}.\\
\begin{figure*}[!t]
\center
\includegraphics[scale=.50]{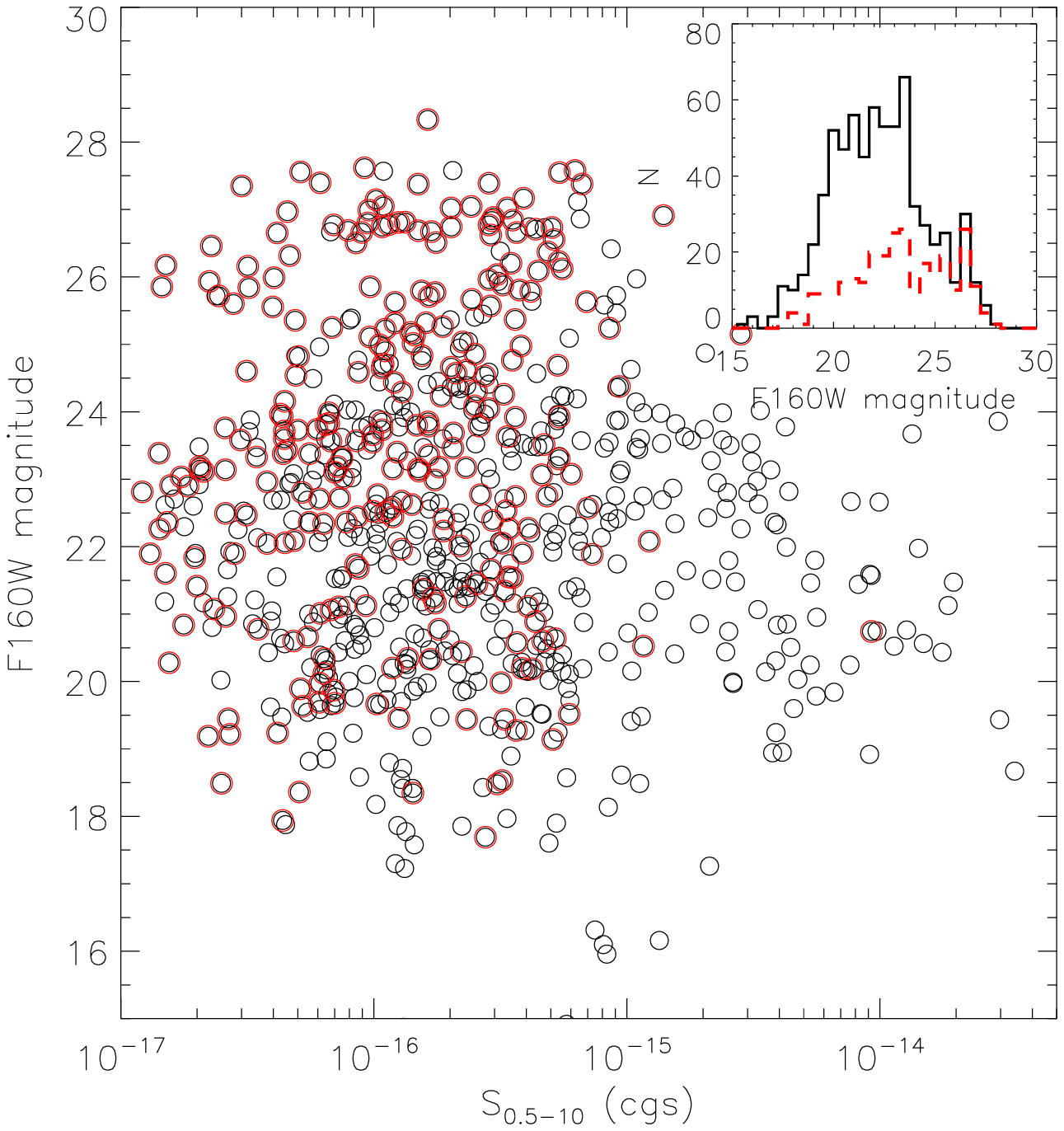}
\includegraphics[scale=.50]{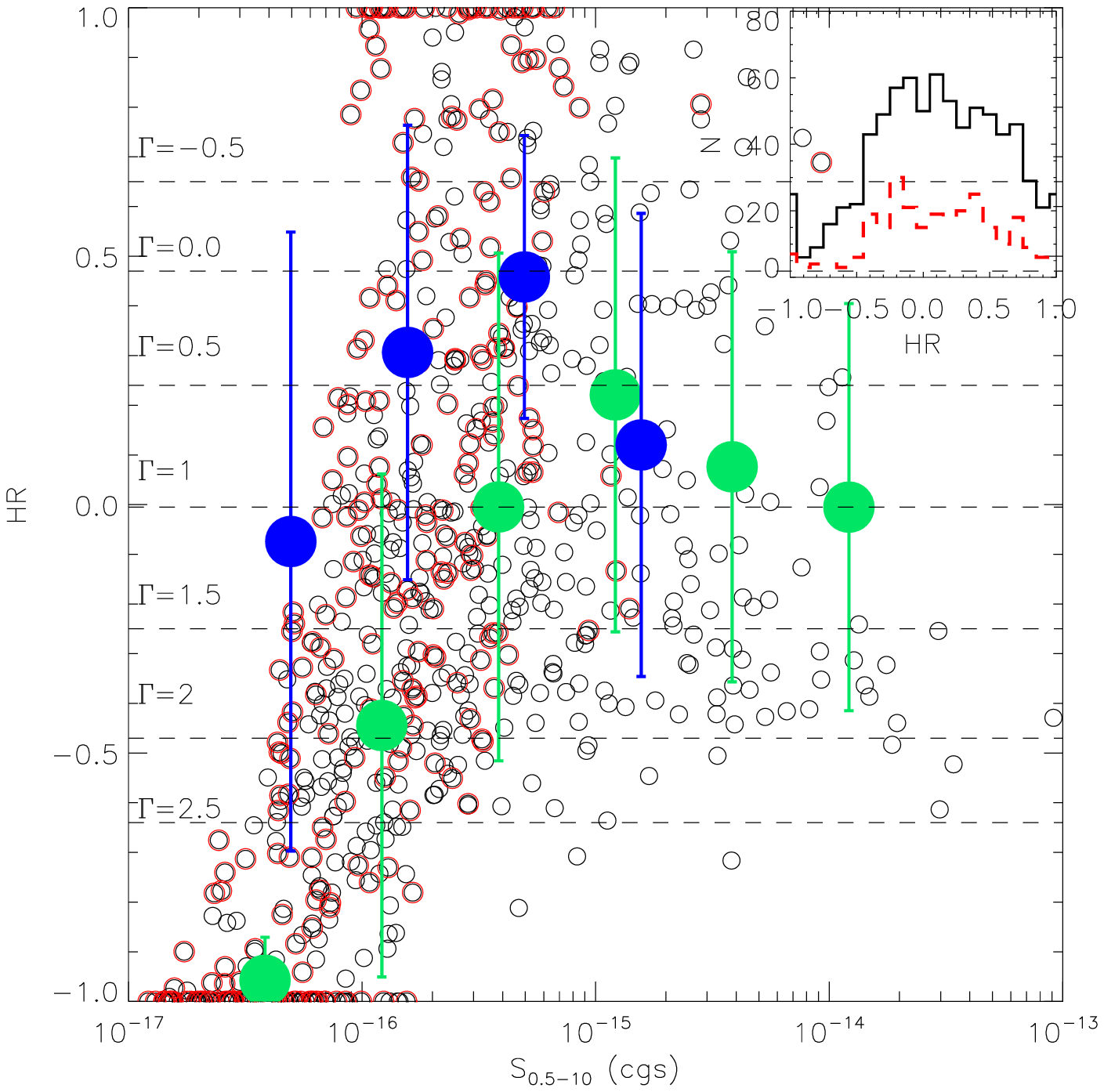}
\includegraphics[scale=.51]{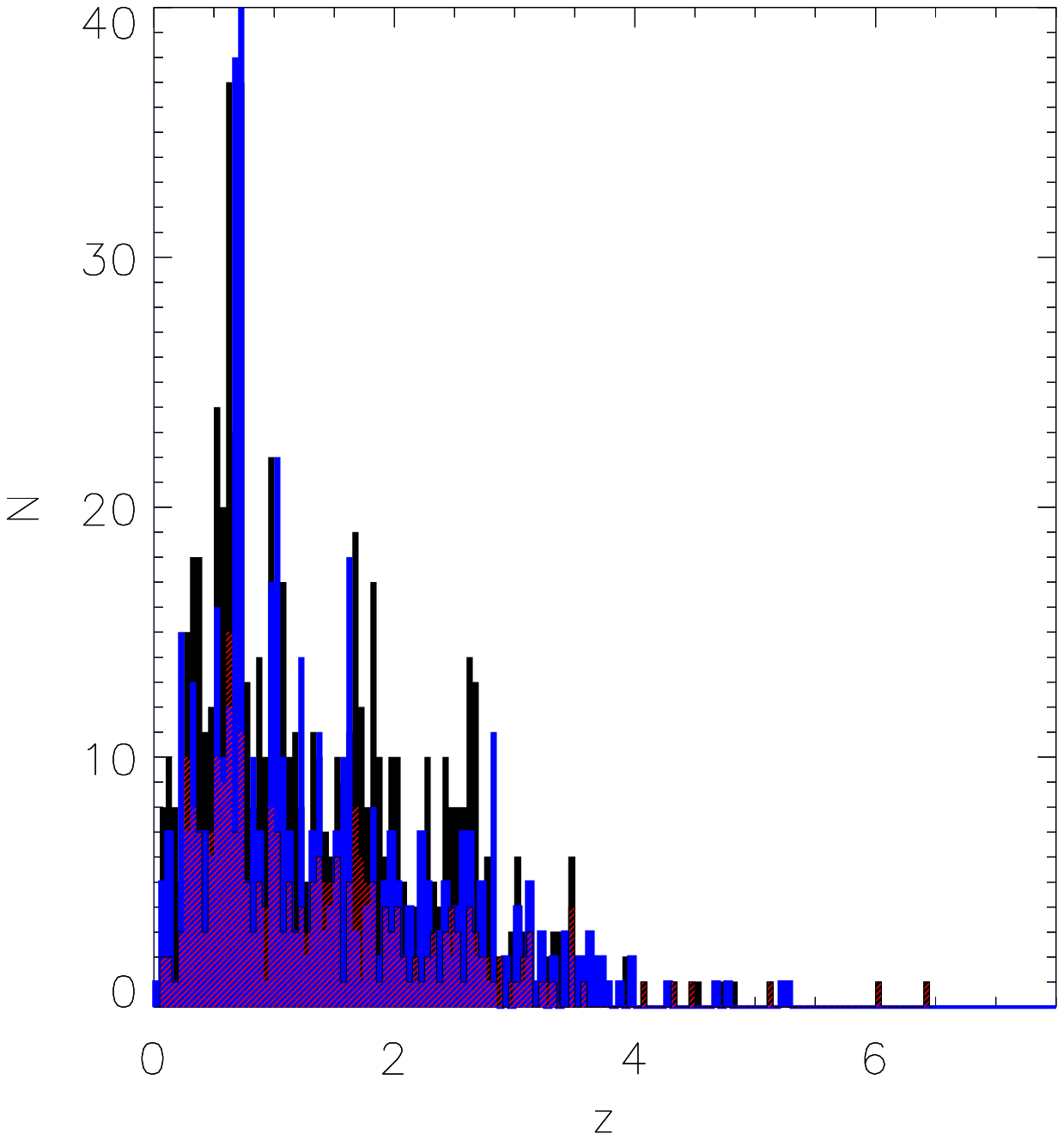}
\includegraphics[scale=.50]{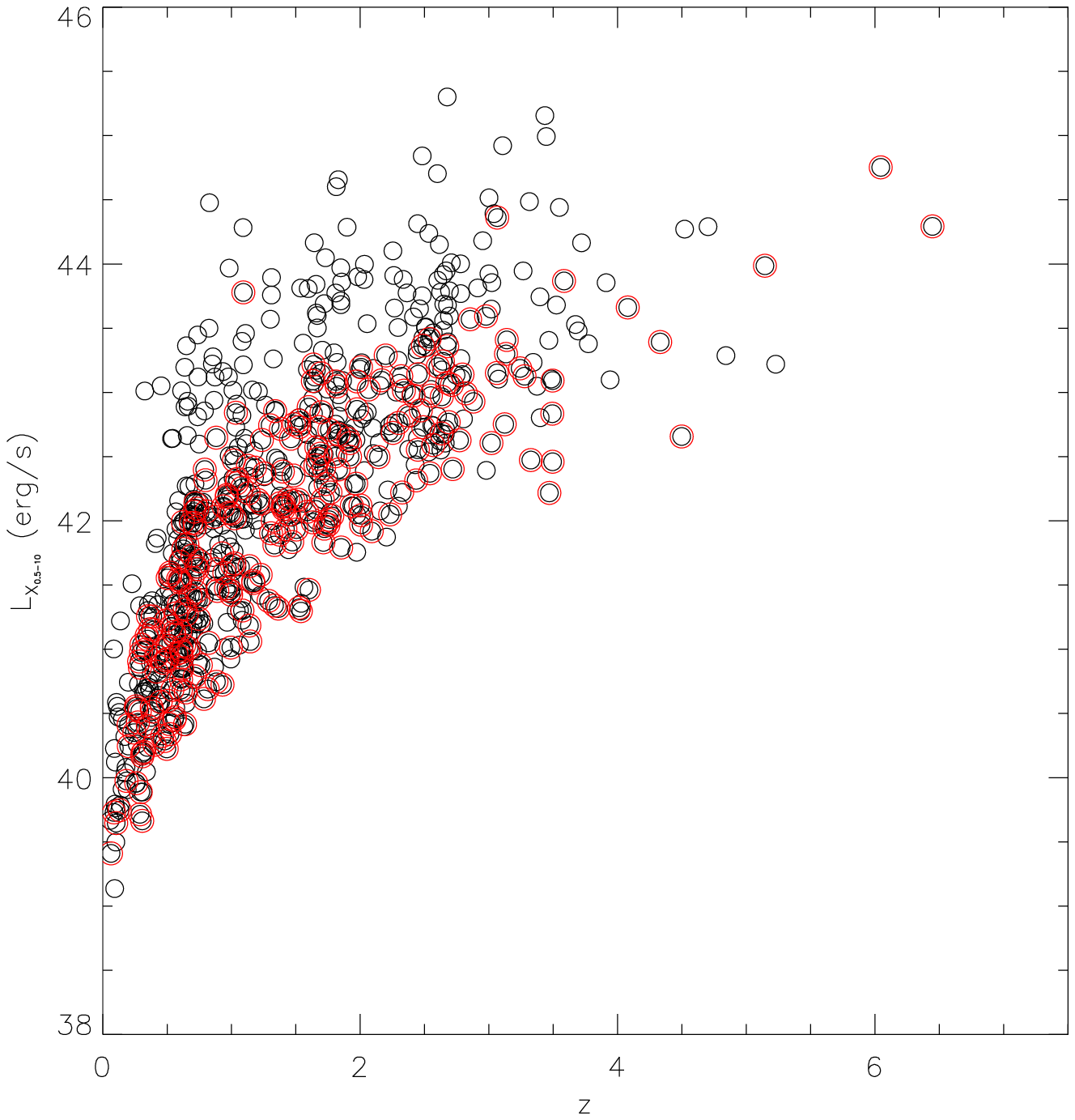}

\caption{ \label{magflux} $Top~left~panel$ The S$_{0.5-10}$ versus the F160W AB magnitude for the whole sample ($black~open~circles$) and for 
new sources  ($red~open~circles$).  The inset show the  F160W AB magnitude  distribution for   for the whole sample ($black~histogram$) and for 
new sources  ($red~histogram$).
$Top~right~panel$  The S$_{0.5-10}$ versus HR  for the whole sample ($black~open~circles$) and for 
new sources  ($red~open~circles$). The $horizontal~dashed~lines$ represent the expected HR for a power-law spectrum with varying spectral index $\Gamma=-0.5--2.5$ from $top~to~bottom$.
The $green~filled~circles$ are the average HR in $\Delta$Log(S)=0.25 flux bin for the whole sample, while  $blue~filled~circles$ are the same but for new sources.
$Lower~left~panel:$ The photo-z distribution for  the whole sample ($black~filled~histogram$) and for 
new sources  ($red~filled~histogram$) compared with the fiducial redshift distribution of X11 ($blue~filled~histogram$)
$Lower~right~panel:$ the photo-z versus L$_{0.5-10}$ for  the whole sample ($black~open~circles$) and for 
new sources  ($red~open~circles$).}
\end{figure*}

\section{General  properties of the X-ray sample} 

Here we present a preliminary overview of the properties of newly detected X-ray sources while, 
a more complete  analysis will be presented in a forthcoming dedicated paper. 
In the upper left panel of Fig. \ref{magflux} left, we show the [0.5-10] keV flux of our detections vs the F160W magnitude of their counterparts 
for the whole sample and for the new detected sources. As expected, the new sources are fainter 
than the whole sample, and also their the brightness distribution of their counterparts is peaked at  fainter magnitudes. 
In particular the whole sample of counterparts has $\langle
m_{F160W}\rangle=$23.1, while 
for the new sources $\langle m_{F160W}\rangle=$24.3.

In the upper-right panel of  Fig. \ref{magflux} we show the X--ray
colors as a function of the [0.5-10] keV flux.  The X--ray color, or
hardness ratio, is defined as HR=$\frac{H-S}{H+S}$ where, H and S are the count rates
in the [2-7] keV and [0.5-2] keV energy bands, respectively. 

The whole sample has an average hardness ratio of
$\sim-0.1$ ($green~points$) corresponding to a power--law spectrum
with photon index $\langle \Gamma\rangle$=1.4 
The {\em new} sources  have a slightly harder average hardness ratio  $\langle HR\rangle\sim$0.0-0.5 ($blue~points$).
This difference, although  marginally significant, suggests
that the {\em new} population may include a large number of obscured
AGN.

The luminosities of the low redshift sources are as low as 10$^{40}$ erg s$^{-1}$ 
(see bottom right panel of Fig. 10) indicating that the bulk of the z$<$1 population
is due to star forming galaxies and low luminosity
obscured AGN. 

An updated catalog of  X--ray sources detected in the CDFS with blind
standard methods was recently assembled (\citet{hsu}), 
merging various catalogs:  X11, \citet{luo},\citet{vir} and \citet{rang}. 
In the CANDELS area 11 sources are not detected by   \citet{hsu} (all
of them in X11). Six out of 11 sources are recovered in our catalog. 
As a consequence, the \citet{hsu} catalog contains 5 sources which were
not detected neither by us nor by X11.
Therefore the total number of bona~fide X-ray sources in the CANDELS GOODS area is 789.
 
Finally, we cross correlated our catalog with the photo-z catalog presented by \citet{santini},
in the lower-left panel of Figure \ref{magflux}  we show the 
photo-z distribution for the {\em new} and {\em old}
X-ray source population compared with that of X11. 
Such a catalog has been derived by computing the weighted average of the Probability 
Distribution Functions  (PDFs) obtained by several teams using galaxy templates.
This could be a problem for some of our sources since their powerful X-ray emission indicates
 AGN activity and therefore a nuclear contamination of the SED. For these sources the photo-z
 may not be reliable however, since \citet{hsu} measure the photo-z by including AGN contamination 
 in the fit we included their photo-z for the sources in common.
We note that the bulk of our {\em new} X-ray sources 
lie at z$\sim$1-3  and, remarkably, we find 9 highly reliable (FLAG\_ASSOC=1)
candidates with photo--z $\geq$4, 2 with Spec-z$\geq$4 (and photo-z$<$4) and another 4 with photo--z $\geq$4
but  FLAG\_ASSOC=2  in the CANDELS catalog. 
We point out that source NID=624, detected on the tail of a bright
off--axis X--ray source, could be a spurious detection. 
Eight of them are in common with the X11 and  \citet{gial} catalogs. 
In Table \ref{tab:highz} we report all the high-z
candidates  and mark those already detected by \citet{gial} and X11. 
The high--z candidates are likely to be AGN with luminosities of the
order of  $10^{43-43.5}$ erg s$^{-1}$. 
 Another source in common  with \citet{gial} is CANDELS ID=29323 (NID=495) with
photo-z=9.73 however, the photo-z of this source  is dominated by artifacts in
 the SED and it is not reported in Table  \ref{tab:highz}.
The high--z candidate sources which are not in common with \citet{gial} and X11 
are in general (except one, NID624)  faint and just above the threshold.
Interestingly, \citet{gial} detects  more (22) candidate z$>$4 X-ray sources; this is apparently
 in contrast with our findings. 
We have then searched our raw catalog, which includes sources down to
$\mathcal{L}=$3, and retrieved 17/22 sources within 2$\arcsec$ from
our X-ray centroid. 
Although found at low threshold we cannot exclude with our method, at a significance level of $\sim$95\%
that these sources (at least in this band) are background fluctuations
at the position of CANDELS galaxy.
Therefore we can explain such a discrepancy with the fact
that the two methods adopt different thresholds and different energy
bands. In fact while we used standard energy ranges \citet{gial} choose the
energy band which could maximize the SNR. 
The analysis of the full {\em Chandra} data set, known as the 7 Ms,
will provide further clues and will be the subject of a future investigation.
 Finally we want to point out that in the catalog of \citet{hsu} none of our 7 high-z candidate
in common with them has a photo-z$>$4. While this requires a deeper investigation. a similar result 
was found by \citet{we} who did not find any z$>$5 source in the same area. 

\begin{table*}
\center
\tablenum{3}
\caption{Candidate z$>$4 X-ray sources based on photo-z. \label{tab:highz}}
\tiny
\begin{tabular}{r r r r r r r r r }
 \hline
  \multicolumn{1}{c}{NID} &
  \multicolumn{1}{c}{PRIOR\_ID} &
  \multicolumn{1}{c}{FLAG\_ASSOC} &
  \multicolumn{1}{c}{$\mathcal{L}_{FULL}$} &
  \multicolumn{1}{c}{FLUX\_FULL} &
  \multicolumn{1}{c}{Spec\_z} &
  \multicolumn{1}{c}{Photo\_z} & 
    \multicolumn{1}{c}{Photo\_z\_H} & 
   \multicolumn{1}{c}{X11} \\
\hline
  624\tablenotemark{b}  & 28496 & 1 & 52.096638 & 1.39$\times10^{-15}$ & -9.0 & 6.045 & -99.0 & -99\\
  306 & 4760 & 2 & 7.5424814 & 6.32$\times10^{-17}$ & -9.0 & 5.78 & -99.0 & -99\\
  295\tablenotemark{a}  & 20765 & 1 & 9.3882885 & 5.73$\times10^{-17}$ & -9.0 & 5.229 & 2.6389 & 521\\
  341 & 25825 & 2 & 5.7868 & 3.48$\times10^{-16}$ & -9.0 & 5.145 & -99.0 & -99\\
  216\tablenotemark{a}  & 19713 & 1 & 9.819359 & 8.01$\times10^{-17}$ & -9.0 & 4.842 & 3.0113 & 392\\
  572\tablenotemark{a}  & 4356 & 1 & 66.05208 & 8.66$\times10^{-16}$ & -9.0 & 4.703 & 1.7139 & 485\\
  599\tablenotemark{a} & 16822 & 1 & 230.46376 & 9.09$\times10^{-16}$ & -9.0 & 4.521 & 3.2327 & 371\\
  59 & 4466 & 1 & 5.5986195 & 2.23$\times10^{-17}$ & -9.0 & 4.498 & -99.0 & -99\\
  510\tablenotemark{a} & 273 & 2 & 8.904563 & 5.91$\times10^{-16}$ & 4.762 & 4.488 & 0.1374 & 403\\
  272 & 14537 & 1 & 5.257827 & 1.33$\times10^{-16}$ & -9.0 & 4.331 & -99.0 & -99\\
  400 & 24833 & 1 & 5.476905 & 2.84$\times10^{-16}$ & -9.0 & 4.079 & -99.0 & -99\\
  575 & 24636 & 2 & 28.0794 & 6.68$\times10^{-16}$ & -9.0 & 4.054 & 3.699 & 602\\
  238 & 4209 & 1 & 8.693058 & 6.63$\times10^{-17}$ & 4.724 & 3.123 & -99.0 & -99\\
  571 & 23382 & 1 & 31.888372 & 7.93$\times10^{-16}$ & 4.379 & 2.294 & 2.4261 & 534\\

\hline
\end{tabular}
  \tablenotetext{1}{Source detected by  Giallongo et al. (2015), Fiore et al. (2012). }
 \tablenotetext{2}{Possibly spurious source on the tail of a bright offaxis X-ray source.}
\end{table*}

\section{Conclusion and summary} 

In this paper we have presented a new X-ray source catalog in the GOODS-S area 
based on the 4 Ms {\em Chandra} CDFS data.  For the first time we produced
a catalog  with both a maximum likelihood PSF fitting technique based on
prior HST galaxy detections as well as an ``a-posteriori'' LR test to confirm the association. 
The method is tested through extensive Monte Carlo ray--tracing 
simulations using  the state of art knowledge of the SFR--L$_X$
scaling relation for star--forming galaxies and  AGN 
population synthesis models for the CXB.

In this paper we developed and tested a technique based on optical/near--infrared
priors to fully exploit the deep observations in the  {\em Chandra}
Deep Field South.
The detection of faint X--ray sources at the limit of the {\em Chandra}
capabilities is based on two approaches. 
Recently, thanks to ultra-deep  multiwavelength survey with HST, like CANDELS
combined with high angular resolution of {\em Chandra} some authors proposed
to  use the entire three-dimensional data-cube (position and energy), and searching for
X--ray counts at the position of high--z galaxies in the GOODS--South
survey assuming that the  angular  resolution of {\em Chandra} is good
enough to locate accurately the position of the X-ray sources.

These approaches complement the previously widely adopted one,  based on 
either wavelets (see e.g X11) or PSF fitting \citep{pucc} of candidates sources 
selected among the most significant background fluctuations. The X--ray selected samples
are then matched with optical/NIR catalogs and the actual counterpart of the  X--ray sources
are assigned using the LR techniques which balances the distance source/counterpart
and the underlying magnitude distribution of the counterparts.

Here  we applied both methods to the X-ray 4Ms data of the Goods-South
region. We first performed a PSF fitting on a sample of HST--WFC3 selected
galaxies  down to a magnitude limit where we reasonably
expect to identify most of the  X--ray source counterparts.
Our results, validated by simulations, indicate that using priors we
can detect objects down to a likelihood threshold that is respect than
in previous works. As a result, we end up increasing the number of 
faint sources detection (Fig. \ref{fig:detml} and
Fig. \ref{fig:scov}).

We also performed a likelihood ratio analysis using well established
techniques to associate the detected sources with the optical
catalog.   The overall result is that through the LR test we can confirm that
among the $\sim$83\% of sources for which a secure match is found, 
at off--axis angles $<$4$\arcmin$, the
counterpart determined by the LR is coincident with the prior in $\sim$99\% of the cases. 
This fraction drops to 92-93\% at larger off--axis angles.  The prior is
the actual counterpart of the identified sources, on average, in 96\%
of the cases.  This observational finding is confirmed by extensive simulations.
For the remaining 17\% (i.e. 90 unsecure, 14 ambiguous, and 7 secure  for which the prior and the LR counterpart
do not match)
we cannot draw any conclusion on the identity of the counterpart.



After fitting the X-ray centroid, the LR test suggests that the use of priors ensures the detection
 of the correct counterpart in at least 87\% of the cases.  For the remaining 13\%, the X-ray centroid
 is significantly displaced from the optical source or the objects are
 at large ($>$ 4$^{\prime}$  off-axis angles. 
 Although it is not always possible to firmly associate HST  and {\em
  Chandra} sources without  running a LR analysis, we note that at least 
  for sources with FLAG\_ASSOC=1 that the counterpart is coincident with the prior in 98\% of the cases if we
consider the inner portion of the field of view $\theta_{offaxis}<$4$\arcmin$.
At larger off--axis angles this fraction drops to 92\%.\\

Our method significantly improves the efficiency in the  detection of 
faint X--ray sources in deep X--ray surveys by taking advantage of the
precise HST positions. Indeed 257 new X--ray sources are discovered down to a flux
of $\sim$1(8)$\times$10$^{-17}$ erg cm$^{-2}$ s$^{-1}$  in the [0.5-2] keV ([0.5-10] keV)  energy
band. \\
The final catalog contains 698 X--ray sources selected in the 
[0.5-7] keV energy range.   552 have a  secure  match with the CANDELS catalog.  
By cross-matching the current catalog with those published in the literature
we were able to estimate that the number of unique X--ray sources in the CANDELS GOODS--S
area sums up to 789. 
Based on photo-z and a few spectro-z, 15 candidates high--redshift  z$>$4 AGN are identified.  Six of
them are in  common with \citet{gial} , the counterpart of 
4 FLAG\_ASSOC=2 sources is ambiguous.  While the discrepancy with previous results \citep{gial} can be
explained as due to different approaches and thresholds adopted, we conclude that the actual number of X--ray selected AGN at z$>$5
remains  very sensitive to the details of the analysis and
ultimately needs deeper and better data to be robustly measured.
Also, since other authors using  different approaches  obtain different results than those reported in the  official catalog \citep[e.g.][]{hsu,we}, 
we want to point out that a discussion of  the photo-z quality included in our catalog is beyond the scope of this paper and it will be discussed elsewhere.

Indeed, the method presented and extensively discussed in this paper may be
obviously extended to many other X--ray surveys where deep 
optical/NIR HST ancillary data are available and  may significantly boost the legacy value 
of these programs. We point out that the most rewarding scientific
return of the method is obtained if it is applied to surveys  designed
to have a constant PSF and a sharp core, like the  COSMOS Legacy
and the UDS {\em Chandra} fields.

\acknowledgements
NC acknowledges the Yale University  YCAA Prize Postdoctoral  Fellowship program.
We acknowledge the contribution of the EC FP7 SPACE project ASTRODEEP (Ref. No: 312725).
ASTRODEEP is a FP7-funded  coordinated and comprehensive program of
 i) algorithm/software development and testing; ii) 
data reduction/release, and iii) scientific data validation/analysis aimed at making Europe the 
world leader in the exploitation of the deepest multi-frequency astronomical survey data.
 NC acknowledges Marcella Brusa and Francesca Civano for discussions about the LR technique.
 NC thanks Roberto Gilli and Cristian Vignali for insightful discussions. 
 NC thanks Mara Salvato for discussions about photo-z quality.
 NC thanks Hermann Brunner for his valuable assistance with {\em cmldetect}. JSD acknowledges
  the support of the European Research Council through the award of an Advanced Grant. 
  NC kindly acknowledges M.M. Lozio for the useful discussions.
  We  especially thank the anonymous referee for the useful comments that 
  significantly improved this paper.


\end{document}